\documentclass[letterpaper]{article} 
\usepackage{aaai24}  
\usepackage{times}  
\usepackage{helvet}  
\usepackage{courier}  
\usepackage[hyphens]{url}  
\usepackage{graphicx} 
\usepackage{natbib}  
\usepackage{caption} 
\usepackage{algorithm}
\usepackage{algorithmic}

\usepackage{booktabs}
\usepackage{subfigure}
\usepackage{multirow}
\usepackage{multicol}
\usepackage{tabularx}
\usepackage{times}

\usepackage{soul}
\usepackage{url}
\usepackage[utf8]{inputenc}
\usepackage{caption}
\usepackage{graphicx}
\usepackage{amsmath}
\usepackage{booktabs}
\usepackage{float}
\usepackage{algorithm}
\usepackage{algorithmic}
\allowdisplaybreaks[4]
\usepackage{cite}
\usepackage{array}
\usepackage{graphicx}
\usepackage{multirow}
\usepackage{pdfpages}
\usepackage{subfigure}
\usepackage{tabularx}
\usepackage{amsmath}
\usepackage{amssymb}
\usepackage{amsthm}
\newtheorem{mydef}{Definition}

\newtheorem{mytheorem}{Theorem}

\usepackage{newfloat}
\usepackage{listings}
\DeclareCaptionStyle{ruled}{labelfont=normalfont,labelsep=colon,strut=off} 
\floatstyle{ruled}
\newfloat{listing}{tb}{lst}{}
\floatname{listing}{Listing}
\title{Joint Learning Neuronal Skeleton and Brain Circuit Topology with Permutation Invariant Encoders for Neuron Classification}
\author {
    Minghui Liao\textsuperscript{\rm 1, 2, 3},
    Guojia Wan\textsuperscript{\rm 1, 2, 3} \thanks{corresponding author},
    Bo Du\textsuperscript{\rm 1, 2, 3~*} 
}
\affiliations {
    \textsuperscript{\rm 1}National Engineering Research Center for Multimedia Software, School of Computer Science, Wuhan University, China\\
    \textsuperscript{\rm 2}Institute of Artificial Intelligence, Wuhan University, China\\
    \textsuperscript{\rm 3}Hubei Key Laboratory of Multimedia and Network Communication Engineering, Wuhan University, China\\

    minghui@whu.edu.cn, guojia@whu.edu.cn, dubo@whu.edu.cn
}

\usepackage{bibentry}

\begin{document}

\maketitle

\begin{abstract}
Determining the types of neurons within a nervous system plays a significant role in the analysis of brain connectomics and the investigation of neurological diseases. However, the efficiency of utilizing anatomical, physiological, or molecular characteristics of neurons is relatively low and costly. With the advancements in electron microscopy imaging and analysis techniques for brain tissue, we are able to obtain whole-brain connectome consisting neuronal high-resolution morphology and connectivity information. However, few models are built based on such data for automated neuron classification. In this paper, we propose NeuNet, a framework that combines morphological information of neurons obtained from skeleton and topological information between neurons obtained from neural circuit. Specifically, NeuNet consists of three components, namely Skeleton Encoder, Connectome Encoder, and Readout Layer. Skeleton Encoder integrates the local information of neurons in a bottom-up manner, with a one-dimensional convolution in neural skeleton's point data; Connectome Encoder uses a graph neural network to capture the topological information of neural circuit; finally, Readout Layer fuses the above two information and outputs classification results. We reprocess and release two new datasets for neuron classification task from volume electron microscopy(VEM) images of human brain cortex and \emph{Drosophila} brain. Experiments on these two datasets demonstrated the effectiveness of our model with accuracy of 0.9169 and 0.9363, respectively. Code and data are available at: \textit{https://github.com/WHUminghui/NeuNet}.
\end{abstract}

\section{Introduction}
Identifying various neuron types in a nervous system is a fundamental task in numerous neuroscience studies. Assigning neuron type is a necessary step to understand the anatomical and functional properties of a brain circuit \cite{armananzas2015towards}. By identifying and classifying various types of neurons, researchers can determine which neurons are most susceptible to causing some neurodegenerative diseases, such as Alzheimer's\cite{buckner2005molecular} and Amyotrophic Lateral Sclerosis (ALS).
However, due to the vast number and minuscule size of neurons, anatomical-based neuron classification methods are exceedingly time-consuming and labor-intensive. 
\begin{figure}[t]
	\centering	
	\includegraphics[width=0.48\textwidth]{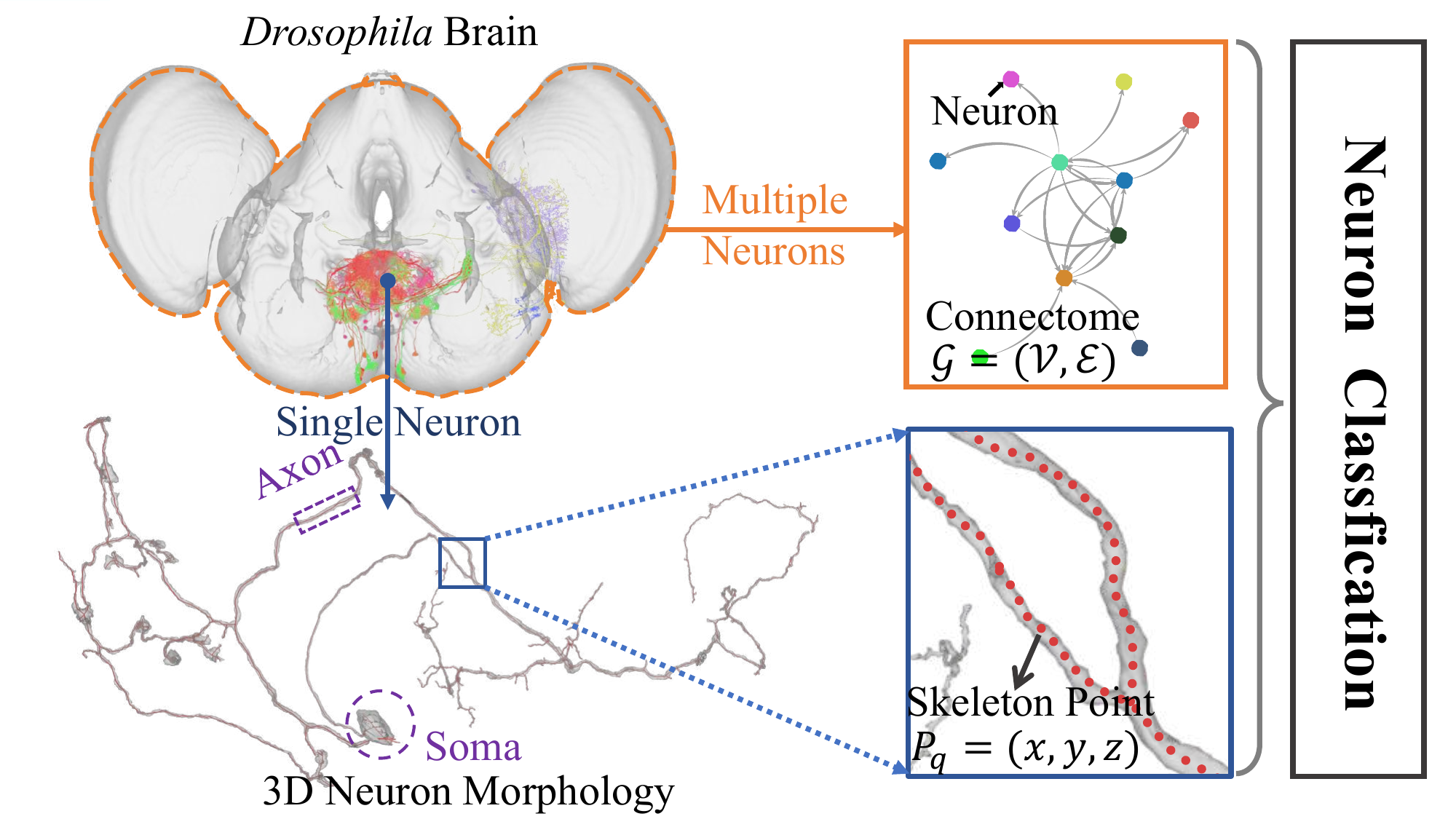}
	\caption{Neuron connectivity (orange) and morphology (blue) in a \emph{Drosophila} (fruit fly) brain. For each neuron, its connectivity can be represented by a graph. Its morphology can be described by a skeleton that is a set of 3D points. Both of them can reveal neuron`s functions and roles.}
	\label{fig1_SC}
\end{figure}
\begin{figure*}[ht]
	\centering	
	\includegraphics[width=0.97\textwidth]{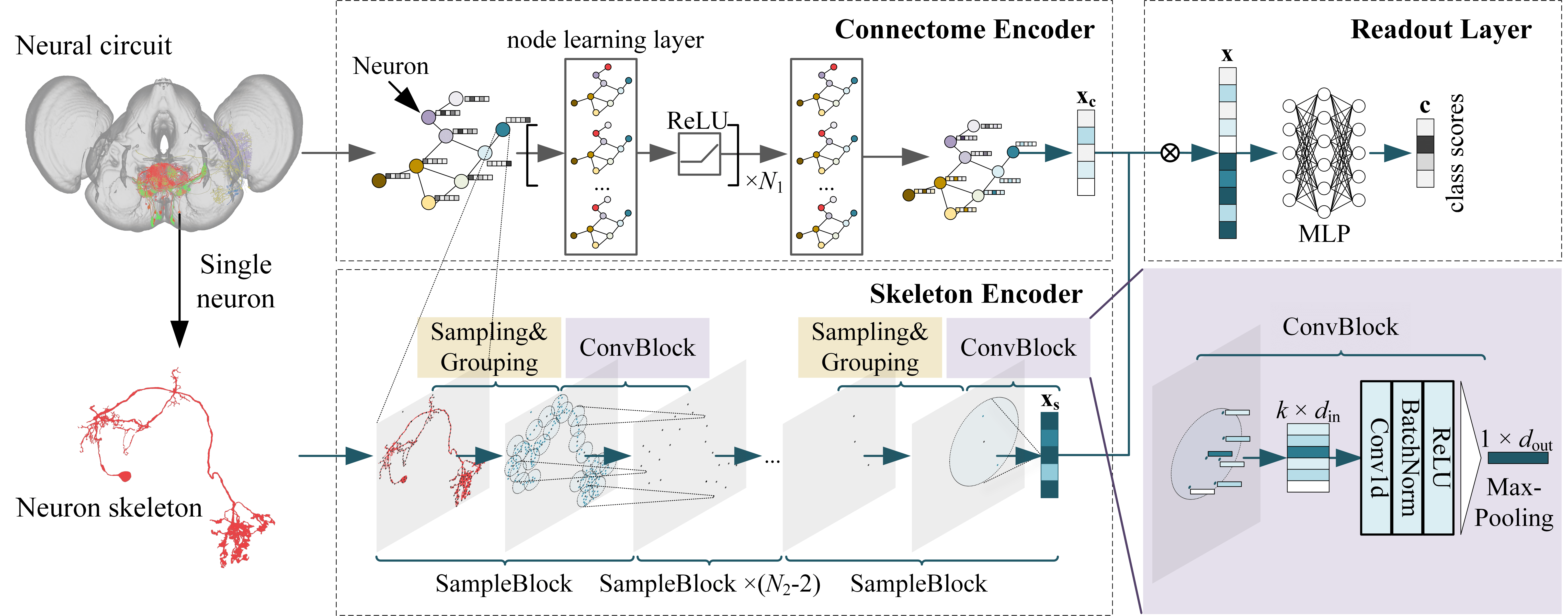}
	\caption{Architecture of proposed NeuNet. On the one hand, the skeleton data of neurons are input into Skeleton Encoder, and Sampling operation samples the points in the skeleton's point set using FPS. Grouping operation divides the attached points into groups with the sampled points as the center, and ConvBlock extracts the features on these group's points and fuses them using max pooling to obtain the group's global features. The above process is repeated until the skeleton's global features are obtained. On the other hand, GNN-based Connectome Encoder mines the topological features of the neural circuit through the information interaction between nodes and neighboring nodes. Finally, Readout Layer fuses the above two features and outputs the classification results with an MLP.}
	\label{fig:Network architecture}
\end{figure*}
Recently, volume electron microscopy (VEM) imaging and analysis techniques \cite{FFN} contribute the availability of whole-brain three-dimensional(3D) reconstruction with cellular resolution, providing a new opportunity to understand the neuronal properties and functions. Specifically, the 3D images of brain tissue generated by VEM contain a wealth of information about neuron size, location, morphology, and connectivity. We show a neuron in Figure \ref{fig1_SC} from \textit{Drosophila} connectome \cite{hemiBrain}. We can observe that the morphology of the complex nerve fibers (including axons and dendrites) that spans multiple brain regions. The red centerline within the fiber voxel is referred as the skeleton of the neuron, which can be depicted as a set filling with 3D points (in XYZ coordinates), $P=\{p_q|q=1,...,m\}$. A large number of such neurons are interconnected, forming the connectome of \emph{Drosophila} brain \cite{hemiBrain}. The connectome can be depicted by a complex network, which is formalized as a graph $\mathcal{G=(V,E)}$. In this network, nodes represent the neurons and edges represent the synaptic connections between them. Apparently, both of these two data exhibit permutation invariance. That is, permuting or rearranging the order of the elements in these data does not change the results or the information conveyed by the data.

The morphology, or shape, of neuron can provide significant insights into its function within the nervous system. Traditional methods based on feature statistics \cite{gillette2015topological,kanari2019objective,morphognn} usually conduct analysis on the optical data. However, the resolution of neuronal morphological data obtained from optical microscopes is relatively low, which cannot provide neuronal connectivity information. Importantly, the connectivity pattern \cite{fulcher2016transcriptional} of neuron can also reveal the properties about its function. 

To date, jointly learning the morphology and connectivity remains unexplored. In this paper, we propose a unified, efficient framework called NeuNet, a neuron classification framework that incorporates morphological information of neuron skeleton and topological information of connectome. NeuNet consists of a skeleton-level feature extractor and a connectome-level feature extractor, which well respect permutation invariance of points and topology connectivity in the input. After extracting the morphological features from neuron skeleton data and the topological information from the connectome, a fusion module combines the above two features to finally obtain the representation of the neuron. The main contributions of NeuNet are as follows:
\begin{itemize}
    \item NeuNet can learn neuronal representations of entire brain circuit in a high-throughput and faster manner and use them for downstream tasks.
    \item Our model incorporates the skeleton's morphological information and the connectome's topological information of neurons, both of which are in a permutation-invariant manner, and it is in line with the actual characteristics of neuron VEM data.
    \item We reprocessed and released two 3D brain reconstruction data. We introduced baseline models from other tasks, and upon comparative evaluation, our model manifested superior classification performance, successfully capturing the latent feature representations of neurons.
\end{itemize}

\section{Related Work}
\label{sec:formatting}

\subsection{Neuron classification} Neurons exhibit diverse properties in terms of connectivity, morphology, and functionality, and the classification of neurons is the cornerstone for exploring the structure and functional mechanisms of brain circuit\cite{wagner2016revealing}. A series of important advances in single‑cell genome‑wide molecular profiling techniques that have developed over the past decade are benefiting cell‑type classification effort~\cite{wagner2016revealing,resendez2016visualization}. Since the functions of neuron are closely related to its morphology, light-microscopy was used to collect whole‑brain catalogs of morphologies~\cite{m1,m2,m3,m4,m5}. Several methods have been proposed for comparing and classifying neurons from light-microscopy~\cite{lm1,lm2,lm3,lm4}.

With the development of electron microscope imaging speed~\cite{e-microscopydevelepoment}, electron microscopy can provide higher resolution and more informative photographs of brain tissue sections. Taking advantage of these innovations, the brain circuit reconstruction have been made to map the connectivity in \emph{Drosophila} optic and antennal lobes~\cite{e-m-1,e-m-2}, the mouse retina, thalamus, and cortex~\cite{e-m-3,e-m-4,e-m-5}. However, there are relatively few works~\cite{blast,learning,morphognn} that use neural networks for feature extraction and classification of the above-described neuron data, and none of which take into account the topological information of brain circuit.

\subsection{Permutation Invariance} The solution of many tasks requires the model to be invariant to some transformations of the data, such as flipping, rotating, and scaling of images, etc., and the well-known convolutional neural network (CNN)~\cite{CNN1,CNN} satisfies this invariance. However, graph data and neuron-level point set data differ from images in that the former is unordered, requiring the model to be invariant to the feed order. Recent approaches have emerged to address the problem of permutation invariance. Graph neural networks (GNNs) are the main methods with permutation invariance on graph data. \cite{semi}proposed graph convolution network (GCN) which motivate the choice of convolutional architecture via a localized first-order approximation of spectral graph convolutions. On the other hand, spatial domain-based approachs~\cite{gcn1,gcn2gat,gcn3} are proposed to update the features of the central node by continuously aggregating the features of disordered neighboring nodes. This make GNNs become mainstream approachs to address feature extraction for graph structure data. Based on the message passing paradigm, an increasing number of GNN models~\cite{gcn1,gcn-2,gcn-3} are proposed to applied graph data with varying properties. Additionally, for the unconnected and unordered point-set dataset, certain methods avoid perturbation of the data point feed order by pooling operations~\cite{DGCNN,pointnet}.

\section{Methodology}
\subsection{Notations and Problem Definition}
\subsubsection{Neuron data.} We designed NeuNet to integrate morphological features about the skeleton and topological features about connectome. A skeleton-level neuron data is represented as a set of points $P=\{p|p=(x,y,z)\}$, where each point $p$ is a corrdinate vector.  

For modeling connectivity, we regard each neuron as a node. Let synapse connections from pre-synaptic cells to post-synaptic cells be represented as edges, where the number of synapses between two neurons is as connection strength. We then constructed a graph to represent the connectome. We denote the graph with $n$ nodes (neurons) by $\mathcal{G=(V,E)}$, where $\mathcal{V}$ represents its node set and $\mathcal{E}$ represents the set of edges. The adjacency matrix of $\mathcal{G}$ is denoted by $\mathbf{A}\in\mathbb{R}^{n\times{n}}$ with its $(i,j)$-th entry $A[i,j]=e$ indicating the strength of the connection between node $i$ and node $j$, and $e$ is numerically equal to the number of synapses connected between neuron $i$ and neuron $j$. 
\subsubsection{Permutation Invariance.}
Our input include a point set and a graph. The point set is inherently unordered. That is, permuting the input order of points within this point set does not affect the skeleton morphology of the neuron. As for the input graph, it is a 2-tuple $\mathcal{(V,E)}$ that consists of two sets: a node set $\mathcal{V}
$ and an edge set $\mathcal{E}$. Likewise, it should follow permutation invariance as well.

Due to its permutation invariance, any learners working with such data should be invariant under transformations. For example, changing the input order of points, nodes and edges should not modify its neuron classification results.

\paragraph{Neuron Classification.} In the task of neuron classification, we are given skeleton-level point set $P$, the connectome level graph $\mathcal{G=(V,E)}$, and the neuron labels denoted by one-hot vector $Y$. Each labeled node $i\in\mathcal{V}$ is associated with a one-hot vector $y_i\subset\{0, 1\}^C$ which encodes its ground-truth class, where $C$ is the number of predefined classes. We needed to learn a function $f_{\theta}(i|(\mathcal{G},P))$ parameterized by $\theta$ which can predict the correct class for a given unlabeled node $i$. 

\subsection{Network Architecture.}
NeuNet consists of three sub-modules (Figure \ref{fig:Network architecture}): \textit{Skeleton Encoder}, \textit{Connectome Encoder}, and \textit{Readout Layer}. Skeleton Encoder is used to extract the morphological information of neuron; Connectome Encoder is used to extract the topological information of the neural circuit; Readout Layer fuses the skeleton-level and connectome-level representation, and output a $1\times C$ classification score vector with a Multilayer Perceptron(MLP).

\subsection{Skeleton Encoder}
Neurons possess intricate branching patterns, encompassing fundamental local structures such as axons, dendrites, and curvatures. These complex branching morphologies often provide insights into the classification and functional roles of neuron \cite{neuromorpho}. However, inferring the functions and classifications of neurons directly from their 3D structures remains a challenging task. Therefore, there arises a necessity to develop morphology feature extractor for neuron morphology, which is able to obtain global representation of the neuron skeleton concerning its local structure.

Considering that neuron skeleton data is an unordered point set, we designed  Skeleton Encoder in bottom-up manner consisting of multiple SampleBlocks, which can extract consistent feature representations from varying input orders of skeleton data pertaining to the same neuron.
 The SampleBlock abstraction level is made of three key operations (Figure \ref{fig:Network architecture}): \emph{Sampling}, \emph{Grouping}, and \emph{Conv1D}.

\subsubsection{Sampling.}The key component of CNN is the convolution kernel, which extracts local features of an image by performing Hadamard product with the certain region (kernel size) of pixels, and acquires a larger range of semantic features by multiple convolutional operations. Drawing inspiration from this concept, we aim to construct local feature extractor working on different regions of a neuron skeleton. Initially, it is imperative to determine the regions with which this feature extractor is intended to interact. Given input points $\{p_1,p_2,..., p_m\}$, we employ iterative farthest point sampling(FPS, shown in Appendix.A)~\cite{fps} to select a subset of points $\{p_{i_1},p_{i_2},..., p_{i_s}\}$ as the centroids of these regions. This selection ensures that the sampled points are the farthest apart(in European distance in this study). In contrast to random sampling, FPS can offers a wider coverage, thereby establishing a more comprehensive receptive field among the given points. This facilitates the acquisition of global morphological information from the input points.

\paragraph{Grouping.} The subsequent task involves delineating the respective extent of these regions around the central points. For a given set of points $P=\{p_1,p_2,..., p_m\}$ and points $S=\{p_{i_1},p_{i_2},..., p_{i_s}\}$ selected by Sampling, we take $S$ as the centroids and find $k$ points in $P$ that are most adjacent to the centroid respectively to form $s$ group. Considering that the density and distribution of points in $P$ are not uniform, we use ball query\cite{pointnet++} to find $k$ points that are within a radius of the centroid point to guarantee a fixed regional scale. The input to the Grouping operator is a point set of size $m$, and the output are groups of point sets of size $s\cdot k $, where each group corresponds to a local region and $k$ is the number of points in the neighborhood of centroid points.

\paragraph{Conv1D.} 

For the points of the groups generated by Grouping operation, they retain an inherent unordered nature. In order to extract consistent features from groups' points irrespective of the order of input points, we employ one-dimensional convolution for capturing local morphology features within the groups. The input skeleton feature $\mathbf{X}_{\rm in} \in \mathbb{R}^{m\times d_\mathrm{in}}$ is fed into Conv1D and is transformed into the next latent skeleton feature $\mathbf{X}_{\mathrm{out}}\in \mathbb{R}^{m \times d_\mathrm{out}}$ as
    \begin{align}
    \mathbf{X}_{\mathrm{out}}=\mathrm{ReLU}(\mathbf{W}\star \mathbf{X}_{\rm in}+\mathbf{b}),
    \label{conv1d}
    \end{align}
 where $\mathbf{W}\in \mathbb{R}^{\kappa \times d_{\mathrm{in}}\times d_{\mathrm{out}}}$ is the kernel and $\kappa$ is the kernel size. $\mathbf{b}$ denotes bias. $\star$ is the valid matrix multiplication. When the initial input is the point set $P$, the corresponding skeleton feature is $\mathbf{X}_{xyz}\in \mathbb{R}^{m\times 3}$. Otherwise, $d_{\mathrm{in}}$ and $d_{\mathrm{out}}$ are intermediate latent feature dimensions as hyperparameters. $m$ is the number of points. For the $j$-th Conv1D layer, $m^{(j)}$ is the number of points. When applying sampling and grouping operations, $m^{(j-1)}$ of the last layer becomes $m^{(j)}=s\cdot k$.
 
 Here, we set the kernel with $\kappa=1$ and stride=1. In this way, Eq. (\ref{conv1d}) becomes a feature learner that is invariant to permutation of a point set along the `$m$' axis. We provide the relevant theoretical proof in Appendix.B.

After each Conv1D layer, we leverage MaxPooling as a downsampling operation and reduce the spatial dimensions. Subsequently, by stacking  $N_2$ \{Sampling, Grouping, ConvBlock, MaxPooling\} layers, the input $\mathbf{X}_{xyz}\in \mathbb{R}^{|P|\times 3}$ are projecting into a morphology representation $\mathbf{x}_{s}\in \mathbb{R}^{1 \times d}$ layer by layer.

\begin{figure*}[h]
	\centering	
	\includegraphics[width=0.75\textwidth]{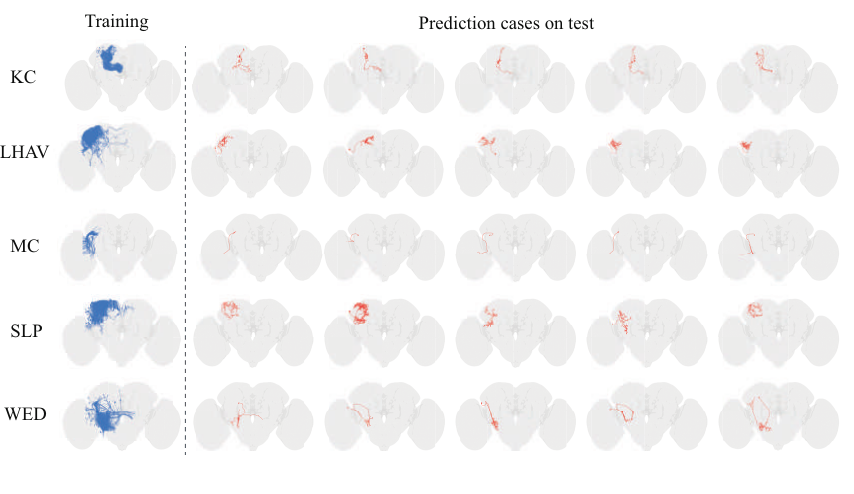}
	\caption{HemiBrain neuron skeletons of KC (Kenyon Cells), LHAV(Lateral Horn Anterior Ventral cell cluster), MC(Medulla Columnar), SLP(Superior lateral protocerebrum), and WED(Wedge). The first column are neurons (blue) from the training set, the rest of the columns are NeuNet's prediction cases on test, and the gray background is \emph{Drosophila} brain. NeuNet learned the morphology of the neurons, including the "L" shaped structure of the KC class.} 
	\label{Skeleton}
\end{figure*}
\begin{table*}[h]
    \centering
    \resizebox{0.75\linewidth}{!}{
    \begin{tabular}{c|c|cccccccccc}
    \toprule
          Method    & Acc   & LC    & KC    & SMP    & AVLP    & SLP    & CL    & PS    & LHAV    & PLP    & LHPV \\
    \midrule
          MLP         & 0.1796    & 0.1763    & 0.1652     & 0.1798      & 0.2032    & 0.1863   & 0.1765     & 0.1956     & 0.1652    & 0.1956    & 0.1698 \\
          PointNet++~\cite{pointnet++}  & 0.8627    & 0.8621    & 0.8962    & 0.75841    & 0.8546    & 0.8654    & 0.8756    & 0.8745    & 0.8522    & 0.8921    & 0.8541 \\
          GCNII~\cite{GCNII}     & 0.5266    & 0.4956    & 0.5361    & 0.5411    & 0.5323     & 0.5362    & 0.5412    & 0.5212    & 0.5421    & 0.4812    & 0.4932 \\
          DGCNN~\cite{DGCNN}     &   0.8560  & 0.8425    & 0.8632    & 0.8410    & 0.8651     & 0.8510     & 0.8710    & 0.8632    & 0.8365    & 0.8863    & 0.8360 \\
          AGNN~\cite{AGNN}      & 0.5327    & 0.5621    & 0.5214    & 0.5412    & 0.5630     & 0.5489   & 0.5321      & 0.5423     & 0.5621    & 0.5421      & 0.5213 \\
          Point-Transformer~\cite{Point_transformer}  & 0.8473  & 0.8516   & 0.8326    & 0.8426    & 0.8321    & 0.8561    & 0.8502    & 0.8632    & 0.8412    & 0.8564    & 0.8214  \\
          NeuNet(ours)    & \textbf{0.9169}    & \textbf{0.9721}    & \textbf{1.0000}    & \textbf{0.9291}    & \textbf{0.9259}    & \textbf{0.9369}    & \textbf{0.9200}    & \textbf{0.9000}    & \textbf{0.8929}    & \textbf{0.9245}    & \textbf{0.8813} \\
    \bottomrule
    \end{tabular}
    }
    \caption{Overall classification accuracy on the HemiBrain dataset and within-class accuracy for 10 of its classes. The corresponding confusion matrix is shown in Appendix.H. Details of the classes we selcted from HemiBrain can be found in Appendix.I.}
    \label{tab:HemiBrain classification}
\end{table*}
\begin{table}[h]
    \centering
    \resizebox{0.99\linewidth}{!}{
    \begin{tabular}{c|c|cccccccccc}
    \toprule
          Method    & Acc   & L1    & L2    & L3    & L4    & L5    & L6 \\
    \midrule
          MLP       & 0.3239    & 0.3364    & 0.3562    & 0.3235     & 0.3102   & 0.3321    & 0.3014 \\
          PointNet++  & 0.6242    & 0.6025    & 0.6821    & 0.6214    & 0.6025    & 0.6412    & 0.6123 \\
          GCNII     & 0.6141    & 0.6231    & 0.6512    & 0.5831    & 0.6451    & 0.5821    & 0.6612 \\
          DGCNN     & 0.6104    & 0.6120    & 0.6521    & 0.6023    & 0.6325    & 0.6230    & 0.5984 \\
          AGNN      & 0.6334    & 0.6231    & 0.6213    & 0.6423    & 0.5921    & 0.6215    & 0.6432  \\
          Point-Transformer & 0.5910    & 0.6012    & 0.5821    & 0.6123    & 0.6215    & 0.6023  & 0.5920 \\
          NeuNet    & \textbf{0.9363}    & \textbf{0.8113}    & \textbf{0.9729}    & \textbf{0.9134}    & \textbf{0.9306}    & \textbf{0.9273}    & \textbf{0.9247} \\
    \bottomrule
    \end{tabular}
    }
    \caption{Overall classification accuracy on the H01 dataset and within-class accuracy on its 6 layers. The corresponding confusion matrix is shown in Appendix.H. L1, L2, ..., L6 are the 6 types of cells at different layers of the human cerebral cortex.}
    \label{tab:H01 classification}
\end{table}
\subsection{Connectome Encoder}

A brain circuit is inherently a graph, with neurons interconnected through synaptic or chemical connections. The connectivity of such graph forms the basis for biological functions and behaviors. For neuron classification, the connectivity pattern can reveal the functions and roles of neurons. To establish a mapping between learning connection patterns and neuron types, we design Connectome Encoder as a graph encoder to learn the topological features of this graph. Connectome Encoder iteratively stacks $N_1$ node learning layers to project graph structure into low-dimensional features. The $l$-th layer of Connectome Encoder is a function $\mathcal{F}:(\mathbf{A},\mathbf{X}^{(l)}\in\mathbb{R}^{|\mathcal{V}|\times d_{(l)}})\to \mathbf{X}^{(l+1)}\in\mathbb{R}^{|\mathcal{V}|\times d_{(l+1)}} $ as
\begin{equation}
\small
\mathbf{X}^{(l+1)} = \mathrm{ReLU}\left(( (1 - \alpha) \mathbf{\hat{P}}\mathbf{X}^{(l)} +
        \alpha \mathbf{X^{(0)}}) ( (1 - \beta) \mathbf{I} + \beta
        \mathbf{\Theta}) \right),
\label{CElayer}
\end{equation}
where $\mathbf{\hat{P}} = \mathbf{\hat{D}}^{-1/2} \mathbf{\hat{A}}\mathbf{\hat{D}}^{-1/2}$ is a degree normalized adjacency matrix with addinng self-loops that $\mathbf{\hat{A}} = \mathbf{A} + \mathbf{I}$. $\mathbf{\hat{D}} $ is a diagonal degree matrix, where $\hat{D}_{ii} = \sum_{j} \hat{A}_{ij}$. $\mathbf{X}^{(0)}\in \mathbb{R}^{|\mathcal{V}|\times d_0}$ is the initial feature matrix from a random initialization. $\mathbf{\Theta} \in \mathbb{R}^{d_{l}\times d_{(l+1)}}$ is a learnable weight matrix.

Here, $\alpha$ models the strength of the initial residual connection, while $\beta$ models the strength of the identity mapping. The final feature $\mathbf{X}_{\mathrm{c}}$  obtained after $N_1$ layers of propagation encapsulates the topological information of the contextual surroundings in which the neuron resides. Then, the row vector $\mathbf{x}_\mathrm{c}\in\mathbf{X}_\mathrm{c}$ represent a neuron's topological low-dimensional feature representation.

Since a graph is a topological structure, its data structure also possesses permutation invariance. This means that permuting the node orders of the graph will not change the learned representation vector of the nodes. Eq. (\ref{CElayer}) also exhibits permutation invariance, and the relevant theoretical proof can be found in Appendix.C.

\subsection{Readout Layer}
After Skeleton Encoder generates feature representations $\mathbf{x}_{s}$ about neuron morphology and Connectome Encoder generates feature representations $\mathbf{x}_{\mathrm{c}}$ about the topological information, we concatenate them to form the final neuron feature representations $\mathbf{x}$, 
\begin{align}
    \mathbf{x}=\mathbf{x}_s\otimes \mathbf{x}_c.
\end{align}
Then, we use an MLP as a classifier to predict neuron label: 
\begin{align}
    \mathbf{c}= \mathrm{MLP}(\mathbf{x}),
\end{align}
where $\mathbf{c}$ denotes the prediction results.

Finally, we use cross entropy as a classification loss to train our model:
\begin{align}
    \mathcal{L}=-\sum^{|C|}_i [y_i\log c_i+(1-y_i)\log (1-c_i)],
\end{align}
where $y_i\in \{0,1\}$ is the groud truth for $i$-th type. $|C|$ is the number of neuron types.
\section{Experiments}
\subsection{Data description}

\subsubsection{HemiBrain.} HemiBrain\cite{hemiBrain} is a dense reconstruction of a portion of the central brain of \emph{Drosophila}. 
This work collected a hemi-brain sample and subdivided it into 20 \textmu m thick slabs using the ‘hot-knife’ ultrathick sectioning procedure~\cite{hot-knife}. Relevant electron images were acquired using FIB-SEM imaging technique~\cite{FIB}. After obtaining these high-resolution photographs, \cite{FFN} and \cite{GAN} were used as segmentation algorithms to determine the regions of neuron bodies and to detect the connecting synapses between neurons based on \cite{synapse_predict}. Finally, the authors grouped the neurons according to their biological properties, location, etc. of each neuron. 

We accessed the raw data through NeuPrint\cite{neuprint} and reprocessed it into 19,384 skeletons(point sets) of neurons which can be classified into 191 different classes(see Appendix.I for details). Regarding these neurons as nodes and the synapses between neurons as connecting edges, these neurons are connected by 3,307,248 edges. More details of the HemiBrain's processing are shown in Appendix.D.
\begin{figure*}[!h]
    \centering
    \subfigure[NeuNet-HemiB]{
        \centering
        \includegraphics[width=0.18\textwidth]{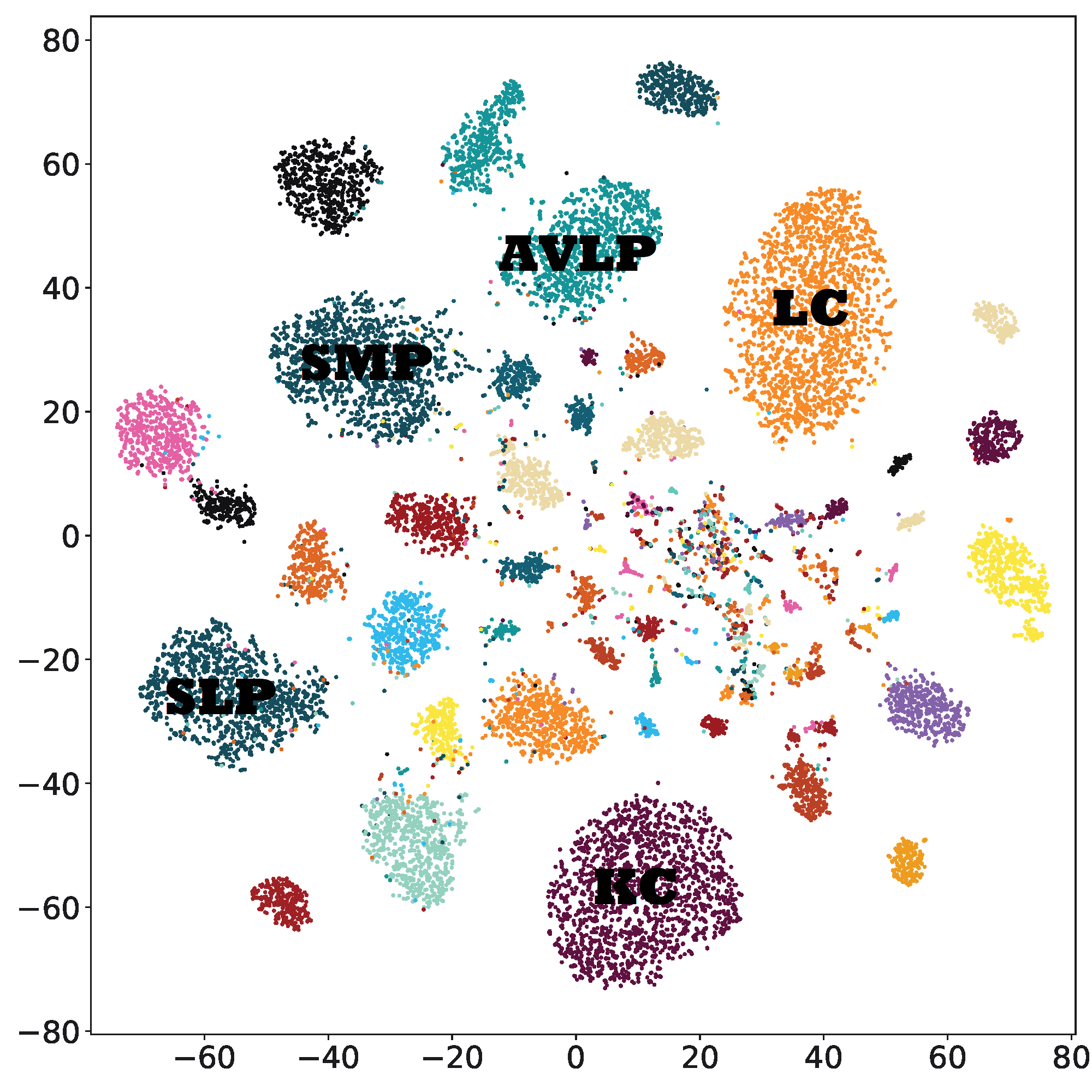}
        \label{HemiBrain-NeuNet}
    }
    \hspace{-5pt}
    \subfigure[PointTrans-HemiB]{
        \centering
        \includegraphics[width=0.18\textwidth]{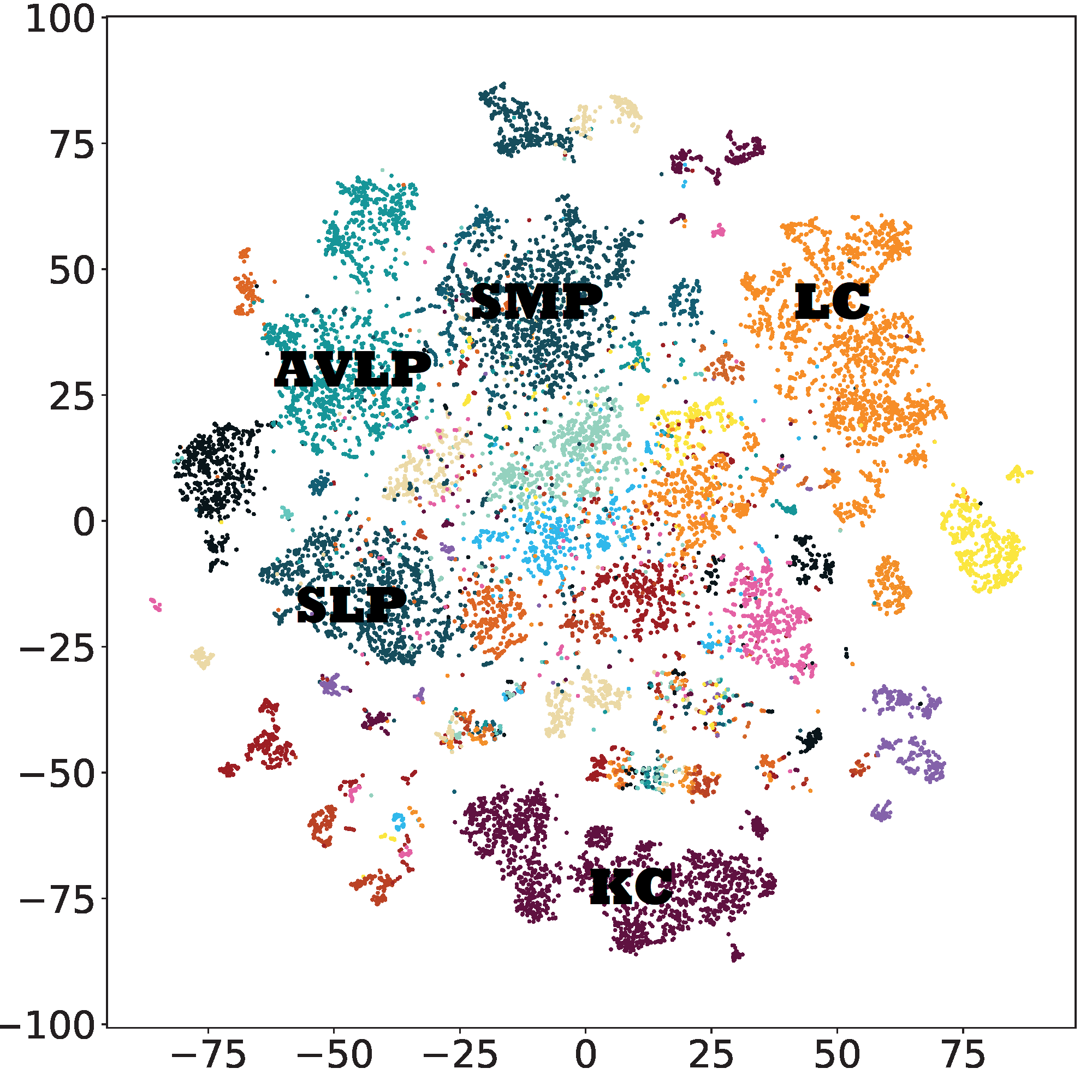}
        \label{HemiBrian-pointTrans}
    }
    \hspace{-5pt}
    \subfigure[DGCNN-HemiB]{
        \centering
        \includegraphics[width=0.18\textwidth]{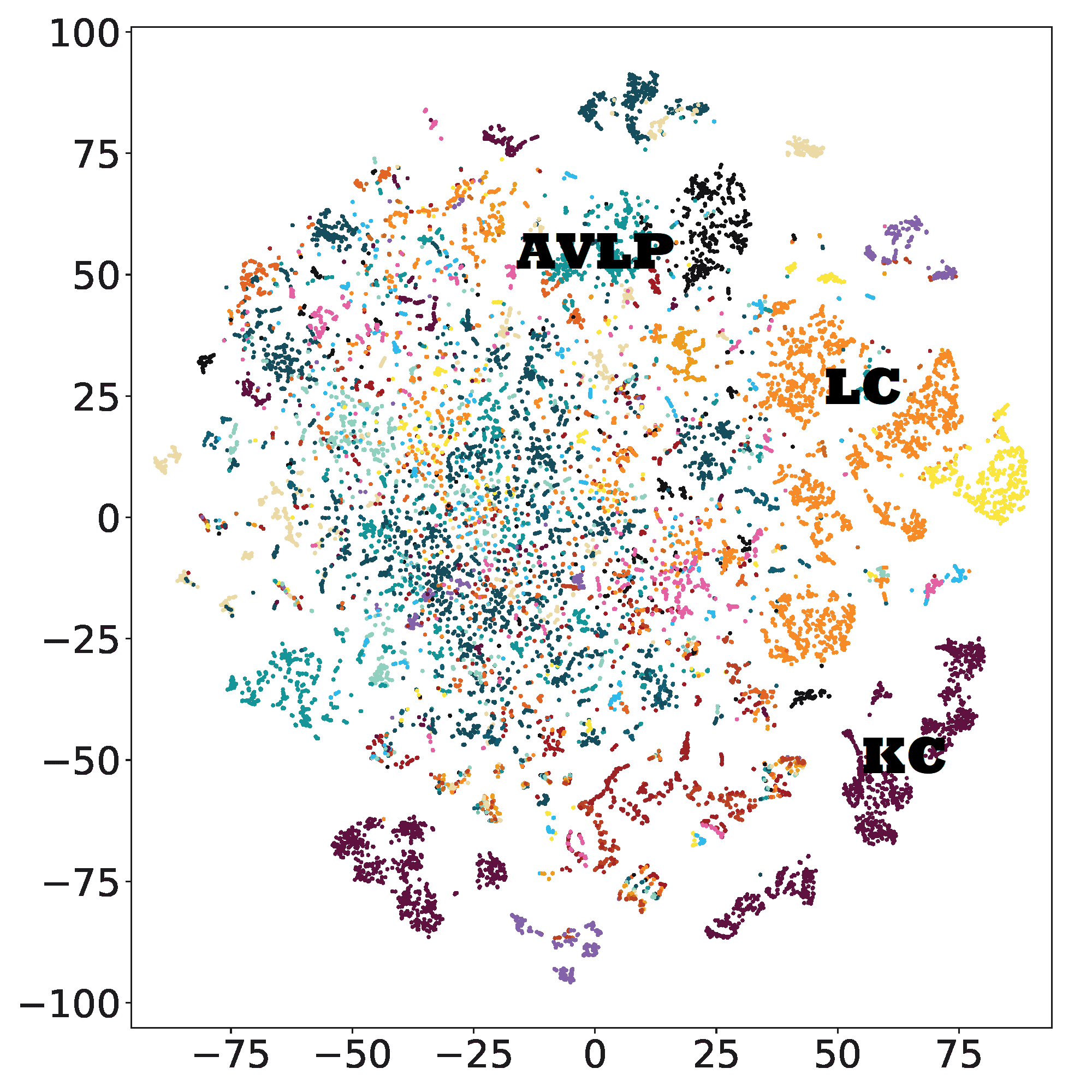}
        \label{HemiBrian-DGCNN}
    }
    \hspace{-5pt}
    \subfigure[PointNet++-HemiB]{
        \centering
        \includegraphics[width=0.18\textwidth]{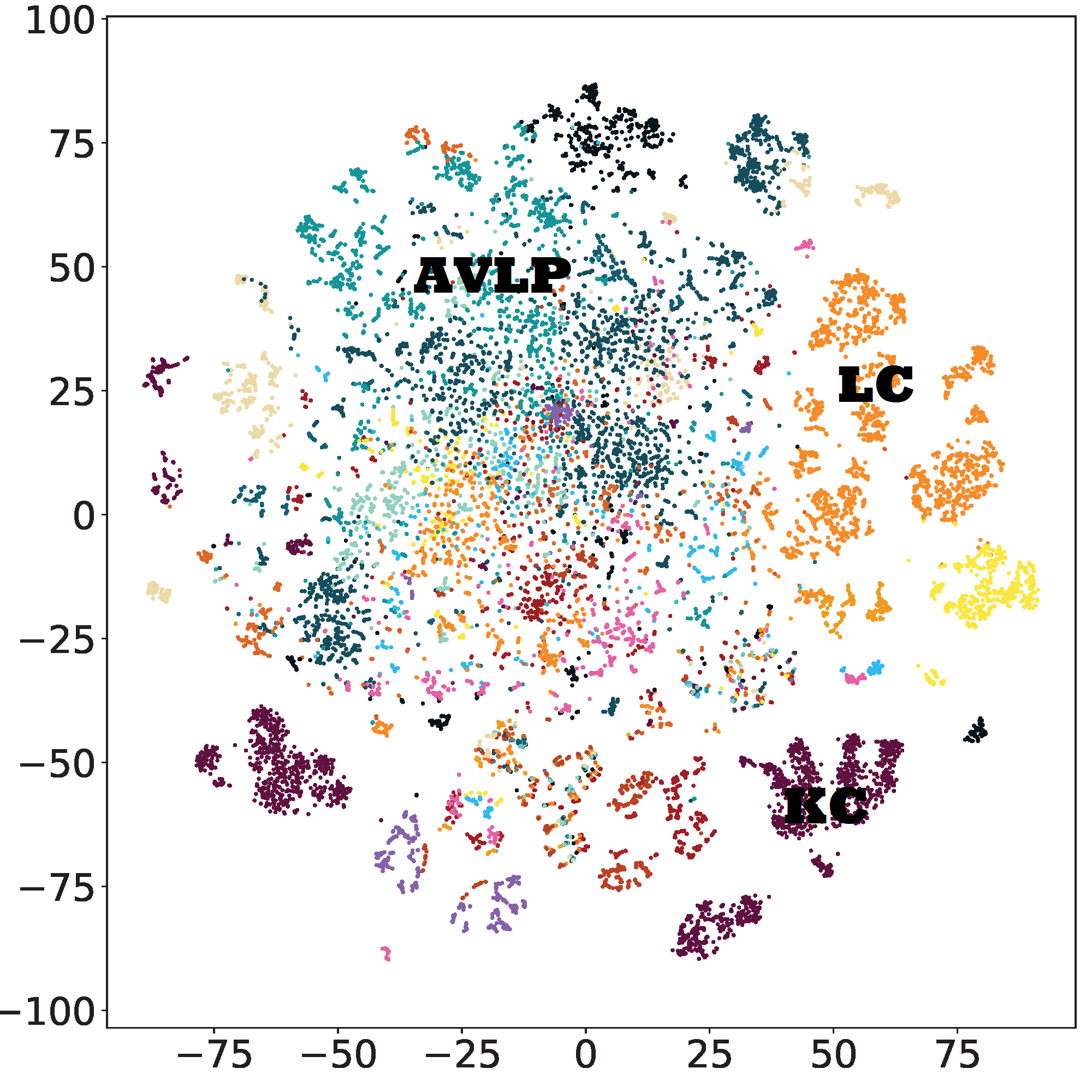}
        \label{HemiBrain-Pointnet++}
    }
    
    \subfigure[NeuNet-H01]{
        \centering
        \includegraphics[width=0.18\textwidth]{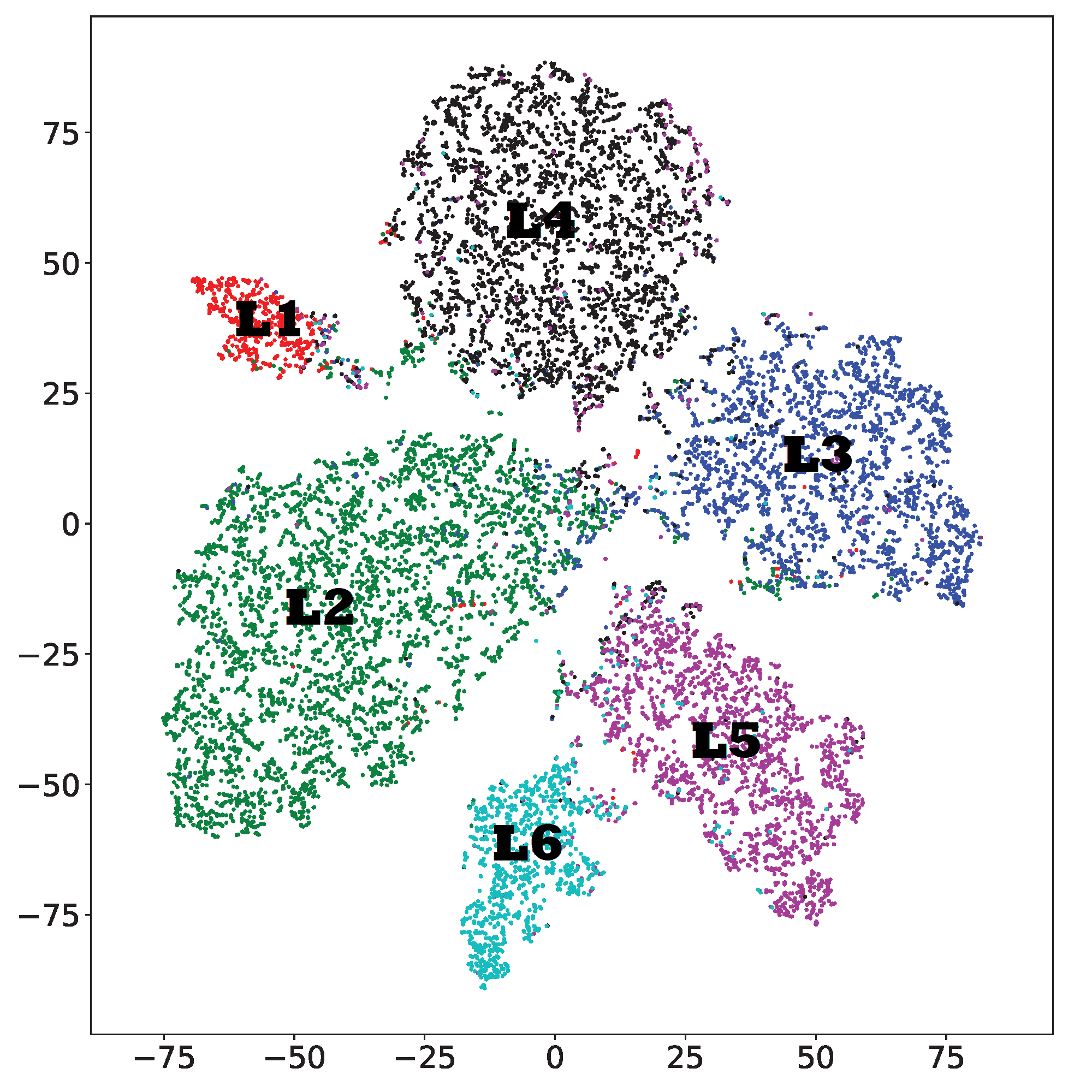}
        \label{H01-NeuNet}
    }
    \hspace{-5pt}
    \subfigure[PointTrans-H01]{
        \centering
        \includegraphics[width=0.18\textwidth]{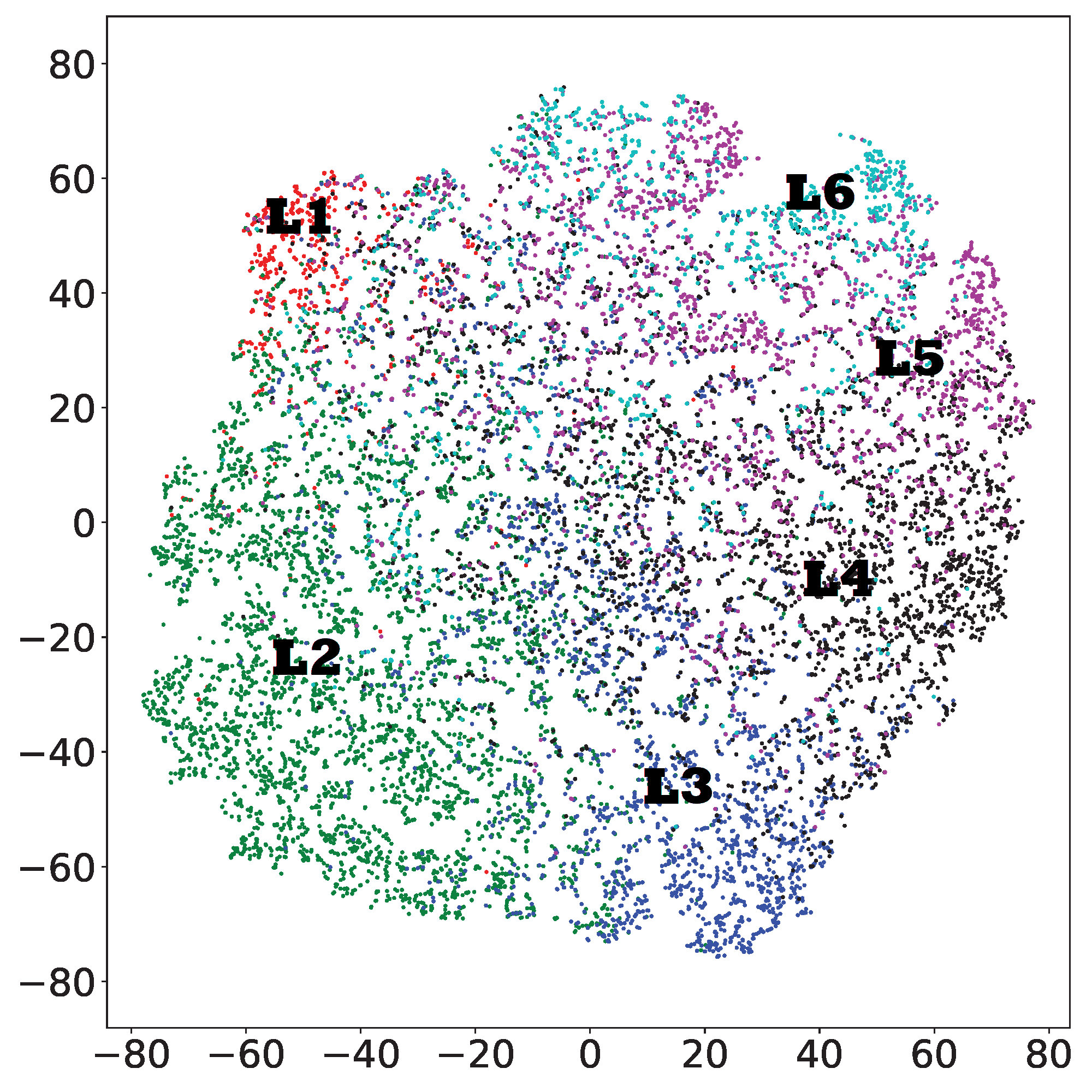}
        \label{H01-pointTrans}
    }
    \hspace{-5pt}
    \subfigure[DGCNN-H01]{
        \centering
        \includegraphics[width=0.18\textwidth]{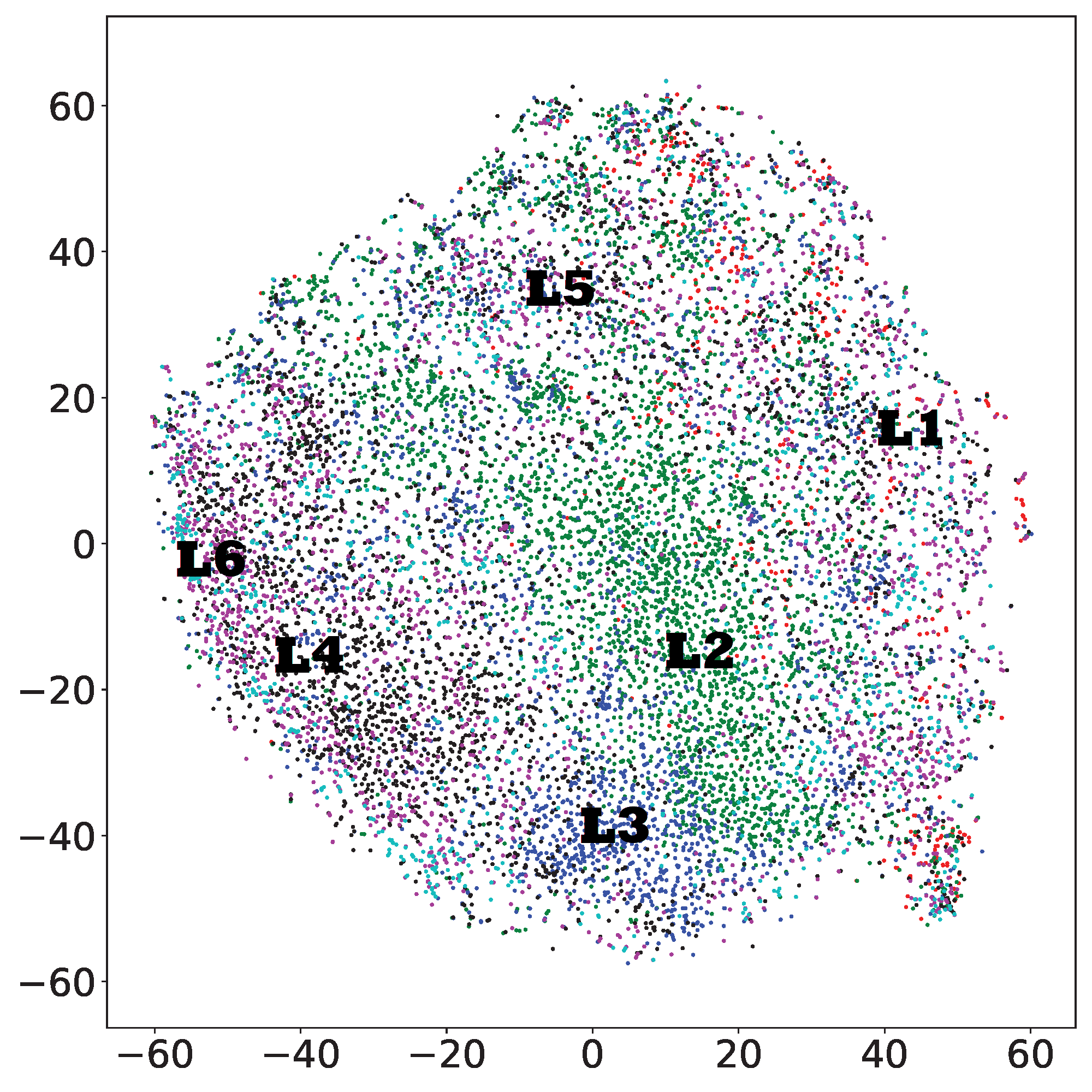}
        \label{H01-DGCNN}
    }
    \hspace{-5pt}
    \subfigure[Pointnet++-H01]{
        \centering
        \includegraphics[width=0.18\textwidth]{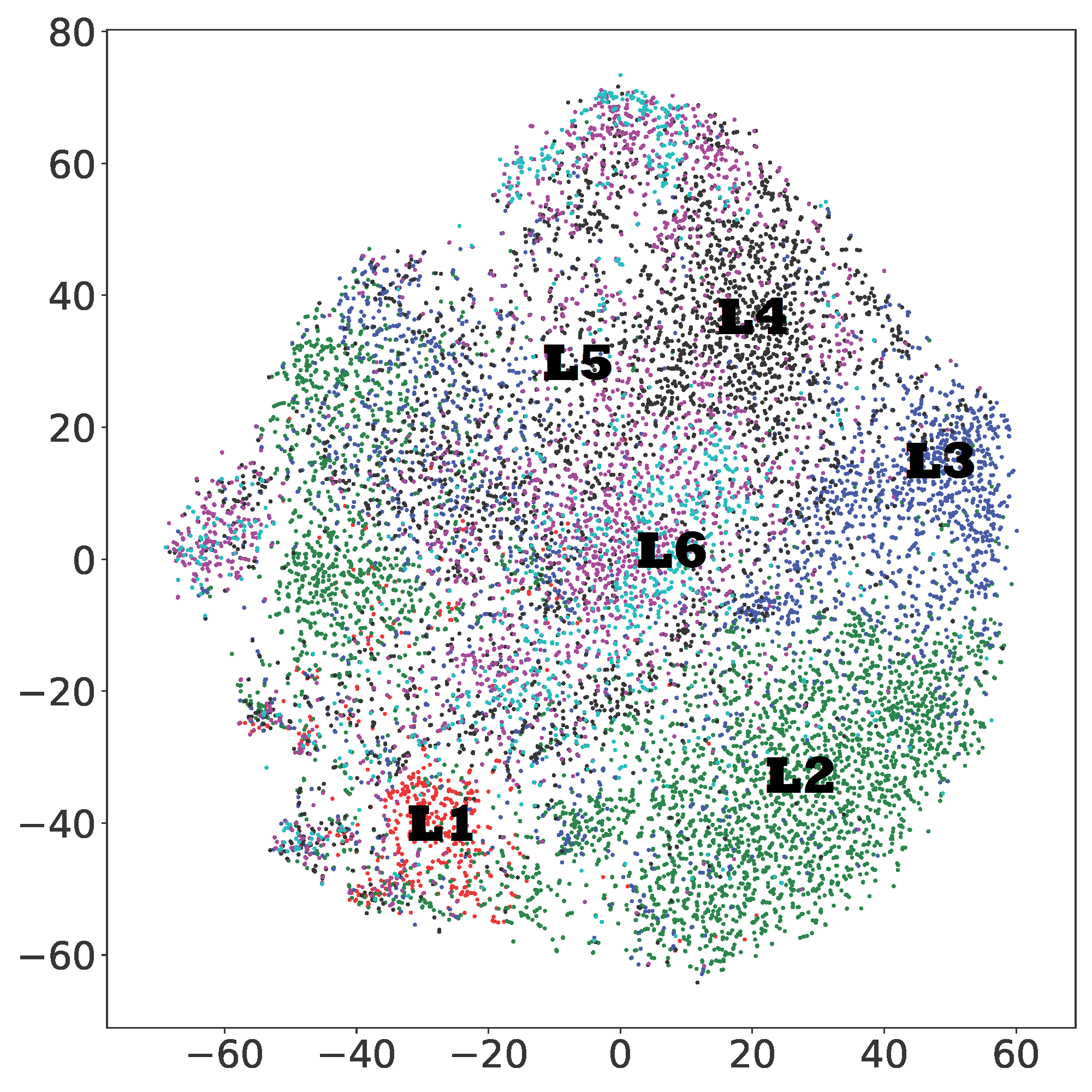}
        \label{H01-Pointnet++}
    }
    \caption{The visualizations using t-SNE~\cite{tSNE} of neuron representation learned by various methods. 
    Where each of color represents a specific class. Since there are up to 191 class of HemiBrain, different classes of point may be given the same color.}
    \label{tsne:Hidden Fea 2d}
\end{figure*}
\subsubsection{H01.} Similar to the HemiBrain dataset, the authors\cite{h01} produced this dataset by taking samples from a human cerebral cortex through steps such as slicing, imaging, segmentation, and synaptic detection. We obtained the raw data and reprocessed it into 15,554 neurons connected by 28,246 edges, which are from 6 layers of the cortex(see Appendix.E for details). 

The neurons of HemiBrain and H01 datasets will be randomly partitioned into training, validation, and test dataset in an 8:1:1 ratio.

\subsection{Implementation Details}
We provide hyper-parameter settings: $\alpha$ in Eq.~(\ref{CElayer}) takes a value of 0.1, and $\beta$ in Eq.~(\ref{CElayer}) takes a value of 0.5(see Appendix.F for related experiments). We set the SampleBlock to 3 layers, and set the node learning layer in Connectome Encoder to 64 layers(see Appendix.G for related experiments). Small changes to the other parameters did not change the results much. 
During the training phase, we concatenate an MLP to Connectome Encoder, and once Connectome Encoder is trained, we proceed to train Skeleton Encoder. For baseline models, we set the parameters same as their original papers.


\subsection{Results}
\subsubsection{Neuron Classification.}
On the HemiBrain and H01 dataset, we compare our method with MLP, PointNet++ \cite{pointnet++}, GCNII \cite{GCNII}, DGCNN \cite{DGCNN}, AGNN \cite{AGNN} and Point-Transformer \cite{Point_transformer}. The classification results are listed in the Table~\ref{tab:HemiBrain classification} and Table~\ref{tab:H01 classification}. The numerical values reported in this study refer to either overall accuracy or within-class accuracy. Overall accuracy represents the proportion of correctly classified samples out of the total number of samples. While within-class accuracy pertains to the proportion of correctly classified samples within a specific category out of the total number of samples in that category. As Table~\ref{tab:HemiBrain classification} suggests, our model outperforms other methods on the HemiBrain dataset. And it is ahead of the second-place model PointNet++ by 6.46\% and ahead of the last-place MLP by 411.41\%. As shown in Table~\ref{tab:H01 classification}, our model also significantly outperforms the other models on the H01 dataset, with an overall accuracy of 48.70\% higher than the second place. 

We draw skeletons of neuron randomly selected from the HemiBrain dataset. As shown in Figure~\ref{Skeleton}, NeuNet is able to learn the local and global structure of neurons' skeletons. For example, NeuNet learns an 'L'-type structure with complex bifurcation structures at both ends in KC; NeuNet recognizes the complex dendritic structure followed by a long axonal structure as a key characteristic in LHAV.

\subsubsection{Representation Visualization.} 
We produce t-SNE \cite{tSNE} visualizations on $\mathbf{x}$ learned by various methods. As shown in Figure~\ref{tsne:Hidden Fea 2d}, our model can transform the complex neuronal data into clustered groups, demonstrating the strong classification ability of the learned representations. However, this ability is not observed in DGCNN and PointNet++. Fuzzy grouping is also observed in Point-Transformer, with the scatter within the groups being more dispersed and overlapping occurring between the groups.

\begin{figure*}[!h]
	\centering	
	\includegraphics[width=0.65\textwidth]{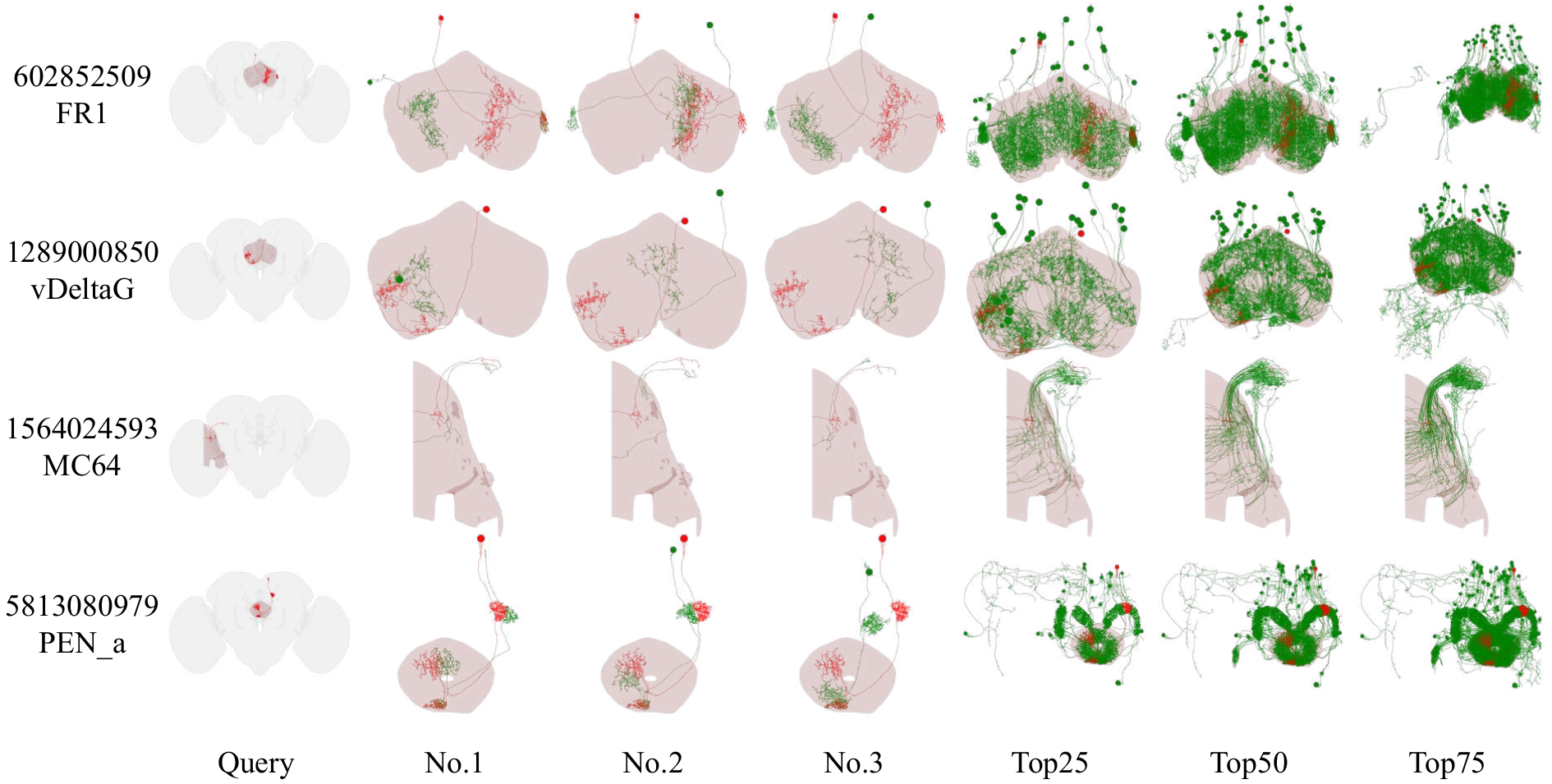}
	\caption{Query neurons(red) and retrieved neurons(green) using NeuNet. The gray-shaded background represents the \emph{Drosophila} brain, and the red shaded background indicates the region with the highest synaptic connectivity for the query neuron. Query neuron IDs and finer-grained categories are provided by~\cite{hemiBrain}. The subcategories of the top 10 retrieved neurons remain consistent with the query neurons(not shown in this figure), without the utilization of finer-grained subcategory information in NeuNet.}
	\label{query}
\end{figure*}
\begin{table*}[!h]
    \centering
    \resizebox{0.75\linewidth}{!}{
    \begin{tabular}{c|c|cccccccccc}
    \toprule
          Method    & Acc   & LC    & KC    & SMP    & AVLP    & SLP    & CL    & PS    & LHAV    & PLP    & LHPV \\
    \midrule
          NeuNet    & \textbf{0.9169}    & 0.9721    & \textbf{1.0000}    & \textbf{0.9291}    & 0.9259    & \textbf{0.9369}    & 0.9200    & \textbf{0.9000}    & 0.8929    & \textbf{0.9245}    & 0.8813 \\
          NeuNet(maxPool)    & 0.9149    & 0.9814    & \textbf{1.0000}    & \textbf{0.9291}    & \textbf{0.9407}    & 0.9279    & \textbf{0.9600}    & 0.8714    & \textbf{0.9821}    & 0.8929    & \textbf{0.9623} \\
          NeuNet(avgPool)    & 0.9135    & \textbf{0.9907}    & \textbf{1.0000}    & 0.9220    & 0.9185    & 0.9279    & 0.9067    & \textbf{0.9000}    & 0.9464    & 0.8571    & 0.9434 \\
          NeuNet(+Skeleton)    & 0.8612    & 0.8512    & 0.8325    & 0.8754    & 0.8695    & 0.8720    & 0.8426    & 0.8365    & 0.8714    & 0.8845    & 0.7965 \\
          NeuNet(+Connnetome)    & 0.5865    & 0.6021    & 0.6231    & 0.5841    & 0.5923    & 0.5814    & 0.5736    & 0.5596    & 0.6021    & 0.6254    & 0.6123 \\
    \bottomrule
    \end{tabular}
    }
    \caption{Classification accuracy of various variants of the NeuNet on the HemiBrain dataset.}
    \label{HemiBrain ablation}
\end{table*}
\begin{table}[!h]
    \centering
    \resizebox{0.99\linewidth}{!}{
    \begin{tabular}{c|c|cccccccccc}
    \toprule
          Method    & Acc   & L1    & L2    & L3    & L4    & L5    & L6 \\
    \midrule
          NeuNet    & 0.9363    & 0.8113    & \textbf{0.9729}    & 0.9134    & \textbf{0.9306}    & 0.9273    & \textbf{0.9247} \\
          NeuNet(maxPool)    & 0.9412    & \textbf{0.9436}    & 0.9422    & \textbf{0.9449}    & 0.9273    & \textbf{0.9462}    & 0.8679 \\
          NeuNet(avgPool)    & \textbf{0.9419}    & 0.9415    & 0.9624    & \textbf{0.9449}    & 0.9273    & \textbf{0.9462}    & 0.8679 \\
          NeuNet-S    & 0.6356    & 0.6541    & 0.6125    & 0.6023    & 0.6152    & 0.6201    & 0.6312 \\
          NeuNet-C    & 0.5123    & 0.5210    & 0.5023    & 0.5421    & 0.5011    & 0.5102    & 0.5321 \\
    \bottomrule
    \end{tabular}
    }
    \caption{Classification accuracy of various variants of NeuNet on the H01 dataset.}
    \label{H01 ablation}
\end{table}
\subsubsection{Neuron Retrieval.}
The representational capacity of NeuNet for high-dimensional neuron data can be applied to neuron retrieval. On the HemiBrain dataset, We use $\mathbf{x}$ to compute the Euclidean distance between all neurons. From the test set, we randomly selecte four neuron representations and identify the nearest neighboring neurons. The experimental results are shown in Figure \ref{query}. It is noteworthy that the subclasses of the top10 neurons are consistent with the query neuron (subclass information of the retrieved neurons is not shown in the figure), despite the fact that subclass information was not utilized during the training phase.

From Figure~\ref{query}, we can observe that the retrieved neurons exhibit a high degree of similarity to the query neuron, and the retrieved neurons share common morphologies and location, indicating the potential presence of functional units within the corresponding regions.

\begin{figure}[!h]
    \centering
    \subfigure[Node scale on HemiBrain]{
        \centering
        \includegraphics[width=0.23\textwidth]{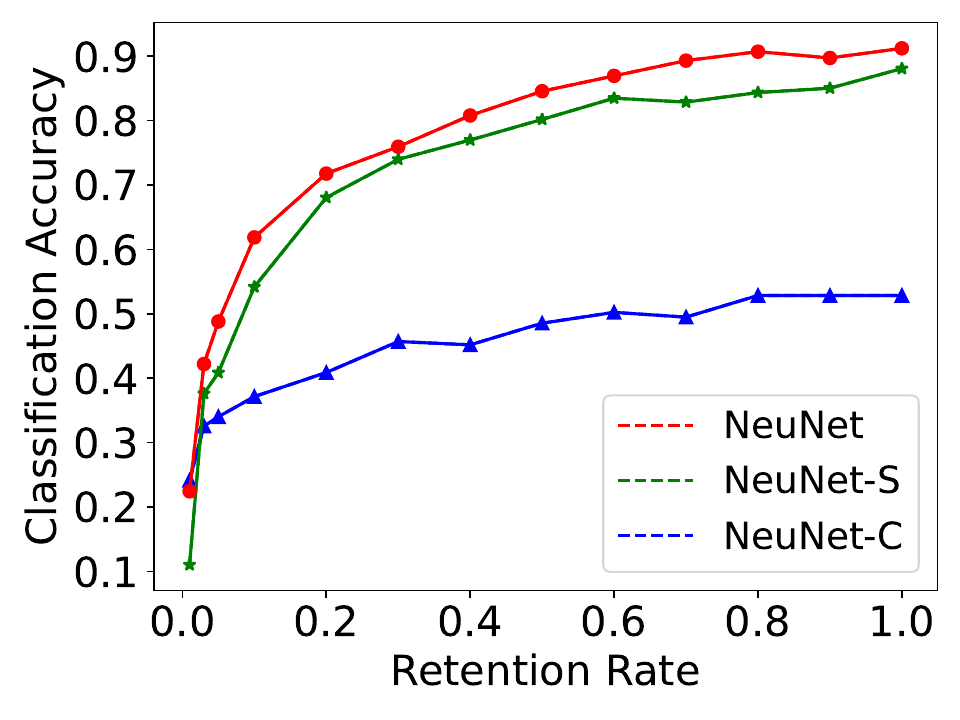}
        \label{node_hemibrain}
    }
    \hspace{-10pt}
    \subfigure[Node scale on H01]{
        \centering
        \includegraphics[width=0.23\textwidth]{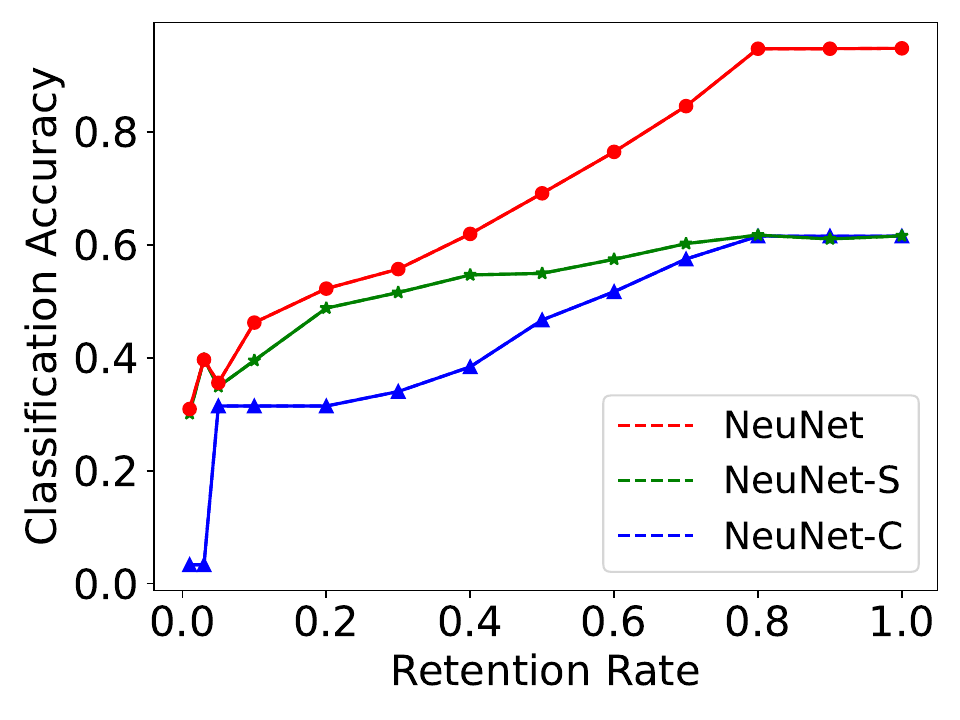}
        \label{node_h01}
    }

    \subfigure[Edge scale on HemiBrain]{
        \centering
        \includegraphics[width=0.23\textwidth]{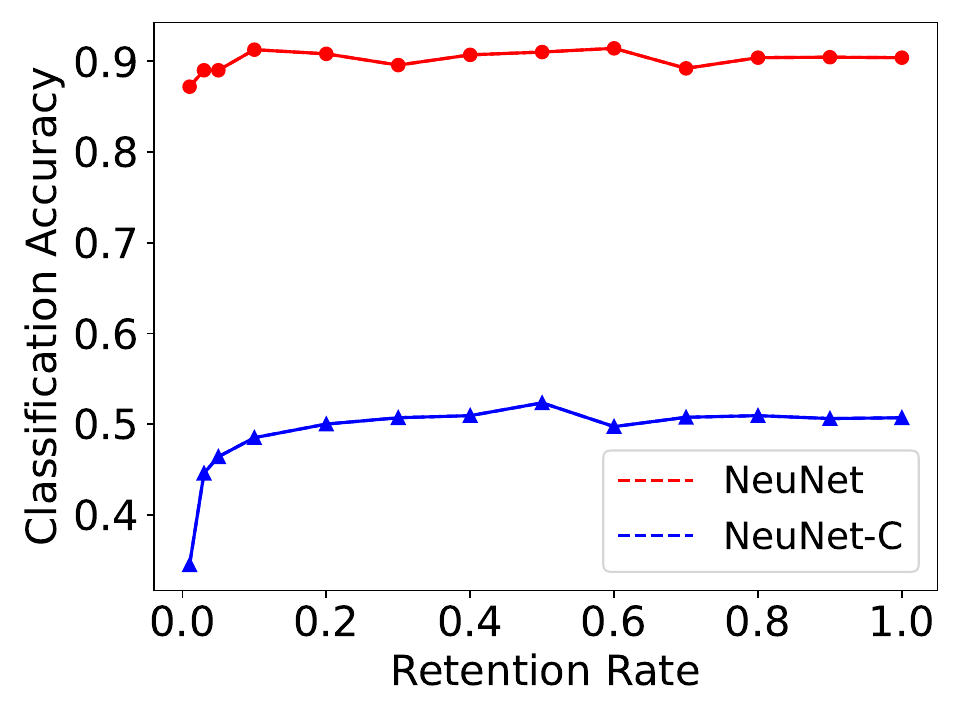}
        \label{edge_hemibrain}
    }
    \hspace{-10pt}
    \subfigure[Edge scale on H01]{
        \centering
        \includegraphics[width=0.23\textwidth]{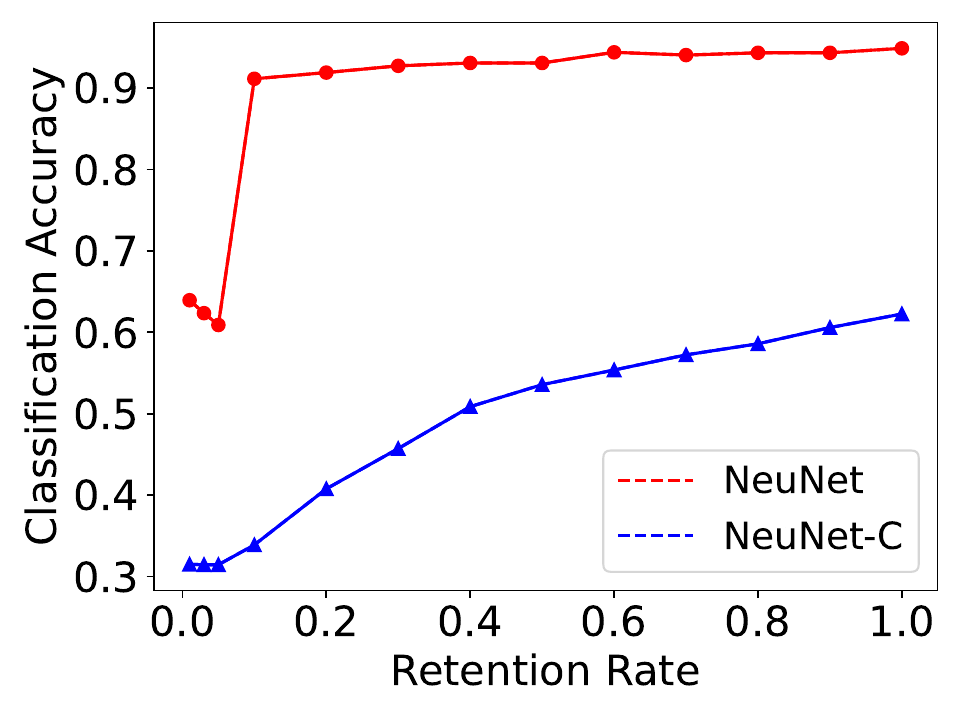}
        \label{edge_h01}
    }    
    
    \caption{The impact of varying the node and edge scale of the training set on classification accuracy.}
    \label{node_edge_scale}
\end{figure}

\subsubsection{Ablation Study.}
We replace the way of fusing features in Readout Layer from concatenation to max pooling and average pooling to construct NeuNet(maxPool) and NeuNet(avgPool), respectively; we discard the brain circuit information and use only the representation generated by Skeleton Encoder to design the NeuNet-S; we remove skeleton information and employ solely the topological information generated by Connectome Encoder to design NeuNet-C. The experimental results of the aforementioned  variants on the HemiBrain and H01 datasets are shown in Table~\ref{HemiBrain ablation} and Table~\ref{H01 ablation}, respectively. It can be seen that the ways of information fusion have little effect. When the NeuNet discards the skeleton information, its effectiveness decreases by 36.03\% and 42.40\% on the HemiBrain and H01 datasets, respectively. Similarly, when NeuNet discards the connectome information, the decreases on the HemiBrain and H01 datasets are 6.07\% and 32.11\%, respectively. This demonstrates that the two types of information complement each other, and NeuNet addresses the issue of the current methods lacking the utilization of brain circuit's topological information.



To investigate the performance of NeuNet under different scales of training samples and connectivity edge quantities, we conduct experiments on the training set while retaining varying proportions of neurons and connectivity edges. 
As shown in Figure~\ref{node_edge_scale}, the model variants achieved 60\% of their optimal performance with only 0.1 times the number of neurons used for training.
Additionally, we can observe that the combination of skeleton and connectome information enhances the model's performance, which is particularly evident on the H01 dataset(Figure~\ref{node_h01}).

\section{Conclusion}
In this paper, we propose a framework called NeuNet specifically for the neuron classification task, which incorporates both the morphological information of neuronal skeleton and the topological information of brain circuit with permutation invariance. NeuNet's Skeleton Encoder mines the morphological information of neuronal sekleton in a bottom-up manner, while NeuNet's Connectome Encoder employs a GNN to extract the topological information of brain circuit. The two information are fused and utilized by the Readout Layer for classification. Furthermore, we reprocess and release two datasets for neuron classification task. On these datasets, NeuNet demonstrate superior performance compared to other baseline models.

\onecolumn

\begin{center}
\LARGE
    Appendix
\end{center}

\section{A. FPS Algorithm}
\begin{algorithm}[]
    \caption{Farthest Point Sampling}
    \begin{flushleft}
    \textbf{Input}: $R$: The set of $n$ points. $K$: The number of sampled points.\\
    \end{flushleft}
   
    \begin{flushleft}
    \textbf{Output}:$S$: The set of sampled points.
    \end{flushleft}
    \begin{algorithmic}[1] 
        \STATE Let $S=\emptyset$;
        \STATE Select point $P_0$ from $R$ randomly; $R$.remove($P_0$); $S$.add($P_0$).
        \FOR{$k=1$; $k<K$; $k++$}
            \STATE $D_R=\emptyset$.
            \FOR{$P_r$ in $R$}
                \STATE $D_S=\emptyset$.
                \FOR{$P_s$ in $S$}
                    \STATE $D_S$.add($Dis_{Eu}(P_s, P_r$)).
                \ENDFOR
                \STATE $D_R$.add(min($D_S$))
            \ENDFOR
            \STATE Set $Dis_{Eu}$($P_S$, $P_R$)==min($D_R$)
            \STATE $R$.remove($P_R$).
            \STATE $S$.add($P_R$).
        \ENDFOR
        \STATE \textbf{return} $S$: The set of sampled points.
    \end{algorithmic}
    \label{FPS}
\end{algorithm}

\section{B. Proof of Translation Invariance on one-dimensional convolution}
\begin{mydef}[Permutation action $\pi$]
A permutation action $\pi$ is a function that acts on any vector, matrix, or tensor defined over the nodes $P$, e.g., $(\mathbf{X}_i)_{i \in P}$, and outputs an equivalent vector, matrix, or tensor with the order of the nodes permuted. We define $\Pi_n$ as the set of all $n!$ such permutation
actions.
\end{mydef}

\begin{mydef}[Permutation invariant function]
Let $\Pi_n$ be the set of all possible permutation action on a set with $n$ elements. A function $f$ is a permutation invariant function iff $\forall \pi \in \Pi_n, f(\pi(\mathbf{X}))=f(\mathbf{X})$, where $\mathbf{X}$ is a vector or matrix with $n$ elements.
\end{mydef}
\begin{mytheorem}
Conv1D(Eq. (1) in main text) is a permutation invariant function,  in the  setting of $\kappa=1$ and stride=1.
\end{mytheorem}

\textbf{Proof:} For a set $\mathbf{X}$, the operation $\pi$ of exchanging order of any element pair $\mathbf{x}_i$ and $\mathbf{x}_j$ has $\forall \pi \in \Pi_n$. So proof that a function is permutation invariant to $\pi$ is equate to proof that this function is permutation invariant to exchang order of elements $\mathbf{x}_i$ and $\mathbf{x}_j$. For $\forall \pi \in \Pi_n$, let the process of applying the Conv1D on $\mathbf{X}$ which has undergone permutation action $\mathbf{\pi}$ is denoted as
\begin{align}
    \mathrm{Con1d(\pi(\mathbf{X}))} & =\mathrm{ReLu}(\mathbf{W} \star \begin{bmatrix}
        x_1 \\
        x_2 \\
        x_j \\
        x_i \\
        \vdots \\
        x_n \\
    \end{bmatrix} + \mathbf{b}) \\
    & = \begin{bmatrix}
        \mathrm{ReLU}(\mathbf{w} \cdot x_1 + b) \\
        \mathrm{ReLU}(\mathbf{w} \cdot x_2 + b) \\
        \mathrm{ReLU}(\mathbf{w} \cdot x_j + b) \\
        \mathrm{ReLU}(\mathbf{w} \cdot x_i + b) \\
        \vdots \\
        \mathrm{ReLU}(\mathbf{w} \cdot x_n + b) \\
    \end{bmatrix} = \pi (\mathrm{Con1D}(\mathbf{X})).
\end{align}
From \textbf{Definition 1}, we have $\pi(\mathbf{X})=\mathbf{X}$ for output value. This implies that
$\mathrm{Conv1D}(\pi(\mathbf{X}))=\pi(\mathrm{Conv1D}(\mathbf{X}))=\mathrm{Conv1D}(\mathbf{X})$.
\begin{flushright}
    Q.E.D
\end{flushright}

\section{C. Proof of Translation Invariance in main text's Eq. (2).}
\begin{mytheorem}
GNN($\cdot$)(Eq. (2) in main text) is a permutation invariant function for input $\mathbf{X}$.
\end{mytheorem}

\textbf{Proof:} From main text's Eq.(2), we can deduce that:
\small{
\begin{align}
            \mathrm{GNN}(\begin{bmatrix}
    x_1 \\
    x_i \\
    x_j \\
    \vdots \\
    x_n \\
\end{bmatrix}) & =  ((1-\alpha)\begin{bmatrix}
    \frac{1}{\sqrt{\hat{d}_1}}    & 0                      & 0                    & \cdots              & 0       \\
    0                       & \frac{1}{\sqrt{\hat{d}_i}}   & 0                    &                     &         \\
    0                       & 0                      & \frac{1}{\sqrt{\hat{d}_j}} &                     &         \\
    \vdots                  &                        &                      & \ddots              & \vdots  \\    
    0                       &                        &                      & \cdots              & \frac{1}{\sqrt{\hat{d}_n}} \\        
\end{bmatrix} \cdot \begin{bmatrix}
    \hat{A}_{11}                  & \hat{A}_{1i}                 & \hat{A}_{1j}               & \cdots              & \hat{A}_{1n}  \\
    \hat{A}_{i1}                  & \hat{A}_{ii}                 & \hat{A}_{ij}               &                     &         \\
    \hat{A}_{j1}                  & \hat{A}_{ji}                 & \hat{A}_{jj}               &                     &         \\
    \vdots                  &                        &                      & \ddots              & \vdots  \\    
    \hat{A}_{n1}                  &                        & \cdots               & \cdots              & \hat{A}_{nn}  \\   
\end{bmatrix} \notag \\
& \cdot \begin{bmatrix}
    \frac{1}{\sqrt{\hat{d}_1}}    & 0                      & 0                    & \cdots              & 0       \\
    0                       & \frac{1}{\sqrt{\hat{d}_i}}   & 0                    &                     &         \\
    0                       & 0                      & \frac{1}{\sqrt{\hat{d}_j}} &                     &         \\
    \vdots                  &                        &                      & \ddots              & \vdots  \\    
    0                       &                        &                      & \cdots              & \frac{1}{\sqrt{\hat{d}_n}} \\        
\end{bmatrix}\cdot\begin{bmatrix}
    x_1 \\
    x_i \\
    x_j \\
    \vdots \\
    x_n \\
\end{bmatrix}+\alpha \begin{bmatrix}
    x_1^{(0)} \\
    x_i^{(0)} \\
    x_j^{(0)} \\
    \vdots \\
    x_n^{(0)} \\
\end{bmatrix})((1-\beta)\mathbf{I}+\beta \mathbf{\Theta}) \\
 & = ((1-\alpha)\begin{bmatrix}
    \frac{\hat{A}_{11}}{\sqrt{\hat{d}_1}}                  & \frac{\hat{A}_{1i}}{\sqrt{\hat{d}_1}}                 & \frac{\hat{A}_{1j}}{\sqrt{\hat{d}_1}}               & \cdots              & \frac{\hat{A}_{1n}}{\sqrt{\hat{d}_1}}  \\
    \frac{\hat{A}_{i1}}{\sqrt{\hat{d}_i}}                   & \frac{\hat{A}_{ii}}{\sqrt{\hat{d}_i}}                 & \frac{\hat{A}_{ij}}{\sqrt{\hat{d}_i}}               &                     &         \\
    \frac{\hat{A}_{j1}}{\sqrt{\hat{d}_j}}                  & \frac{\hat{A}_{ji}}{\sqrt{\hat{d}_j}}                 & \frac{\hat{A}_{jj}}{\sqrt{\hat{d}_j}}               &                     &         \\
    \vdots                                     &                                           &                                         & \ddots              & \vdots  \\    
    \frac{\hat{A}_{n1}}{\sqrt{\hat{d}_n}}                  &                                           & \cdots                                  & \cdots              & \frac{\hat{A}_{nn}}{\sqrt{\hat{d}_n}}  \\   
 \end{bmatrix}\cdot \begin{bmatrix}
    \frac{1}{\sqrt{\hat{d}_1}}    & 0                      & 0                    & \cdots              & 0       \\
    0                       & \frac{1}{\sqrt{\hat{d}_i}}   & 0                    &                     &         \\
    0                       & 0                      & \frac{1}{\sqrt{\hat{d}_j}} &                     &         \\
    \vdots                  &                        &                      & \ddots              & \vdots  \\    
    0                       &                        &                      & \cdots              & \frac{1}{\sqrt{\hat{d}_n}} \\        
\end{bmatrix} \notag \\
& \cdot\begin{bmatrix}
    x_1 \\
    x_i \\
    x_j \\
    \vdots \\
    x_n \\
\end{bmatrix} +\alpha \begin{bmatrix}
    x_1^{(0)} \\
    x_i^{(0)} \\
    x_j^{(0)} \\
    \vdots \\
    x_n^{(0)} \\
\end{bmatrix})((1-\beta)\mathbf{I}+\beta \mathbf{\Theta}) \\
& = (1-\alpha)(\begin{bmatrix}
    \frac{\hat{A}_{11}}{\sqrt{\hat{d}_1}\sqrt{\hat{d}_1}}  & \frac{\hat{A}_{1i}}{\sqrt{\hat{d}_1}\sqrt{\hat{d}_i}}                 & \frac{\hat{A}_{1j}}{\sqrt{\hat{d}_1}\sqrt{\hat{d}_j}}               & \cdots              & \frac{\hat{A}_{1n}}{\sqrt{\hat{d}_1}\sqrt{\hat{d}_n}}  \\
    \frac{\hat{A}_{i1}}{\sqrt{\hat{d}_i}\sqrt{d_1}}  & \frac{\hat{A}_{ii}}{\sqrt{\hat{d}_i}\sqrt{\hat{d}_i}}                 & \frac{\hat{A}_{ij}}{\sqrt{\hat{d}_i}\sqrt{\hat{d}_j}}               &                     &         \\
    \frac{\hat{A}_{j1}}{\sqrt{\hat{d}_j}\sqrt{\hat{d}_1}}  & \frac{\hat{A}_{ji}}{\sqrt{\hat{d}_j}\sqrt{\hat{d}_i}}                 & \frac{\hat{A}_{jj}}{\sqrt{\hat{d}_j}\sqrt{\hat{d}_j}}               &                     &         \\
    \vdots                               &                                           &                                         & \ddots              & \vdots  \\    
    \frac{\hat{A}_{n1}}{\sqrt{\hat{d}_n}\sqrt{\hat{d}_1}}   &                                           & \cdots                                  & \cdots              & \frac{\hat{A}_{nn}}{\sqrt{\hat{d}_n}\sqrt{\hat{d}_n}}  \\   
 \end{bmatrix}\cdot\begin{bmatrix}
    x_1 \\
    x_i \\
    x_j \\
    \vdots \\
    x_n \\
 \end{bmatrix} \notag \\
& + \alpha \begin{bmatrix}
    x_1^{(0)} \\
    x_i^{(0)} \\
    x_j^{(0)} \\
    \vdots \\
    x_n^{(0)} \\
\end{bmatrix})((1-\beta)\mathbf{I}+\beta \mathbf{\Theta}) \\
& = ((1-\alpha)\begin{bmatrix}
    \sum_{k=1}^{n}  \frac{\hat{A}_{1k}}{\sqrt{\hat{d}_1}\sqrt{\hat{d}_k}}x_k \\
    \sum_{k=1}^{n} \frac{\hat{A}_{ik}}{\sqrt{\hat{d}_i}\sqrt{\hat{d}_k}}x_k \\
    \sum_{k=1}^{n} \frac{\hat{A}_{jk}}{\sqrt{\hat{d}_j}\sqrt{\hat{d}_k}}x_k \\
    \vdots \\
    \sum_{k=1}^{n} \frac{\hat{A}_{nk}}{\sqrt{\hat{d}_n}\sqrt{\hat{d}_k}}x_k
\end{bmatrix} + \alpha \begin{bmatrix}
    x_1^{(0)} \\
    x_i^{(0)} \\
    x_j^{(0)} \\
    \vdots \\
    x_n^{(0)} \\
\end{bmatrix})((1-\beta)\mathbf{I}+\beta \mathbf{\Theta}) \\
& = ((1-\alpha)\begin{bmatrix}
    \sum_{k=1}^{n}  \frac{\hat{A}_{1k}}{\sqrt{\hat{d}_1}\sqrt{\hat{d}_k}}x_k  + \alpha x_1^{(0)}\\
    \sum_{k=1}^{n} \frac{\hat{A}_{ik}}{\sqrt{\hat{d}_i}\sqrt{\hat{d}_k}}x_k  + \alpha x_i^{(0)} \\
    \sum_{k=1}^{n} \frac{\hat{A}_{jk}}{\sqrt{\hat{d}_j}\sqrt{\hat{d}_k}}x_k  + \alpha x_j^{(0)} \\
    \vdots \\
    \sum_{k=1}^{n} \frac{\hat{A}_{nk}}{\sqrt{\hat{d}_n}\sqrt{\hat{d}_k}}x_k  + \alpha x_n^{(0)}
\end{bmatrix})((1-\beta)\mathbf{I}+\beta \mathbf{\Theta}) \\
& = \begin{bmatrix}
    ((1-\alpha)\sum_{k=1}^{n} \frac{\hat{A}_{1k}}{\sqrt{\hat{d}_1}\sqrt{\hat{d}_k}}x_k  + \alpha x_1^{(0)})((1-\beta)+\beta \mathbf{\theta}) \\
    ((1-\alpha)\sum_{k=1}^{n} \frac{\hat{A}_{ik}}{\sqrt{\hat{d}_i}\sqrt{\hat{d}_k}}x_k  + \alpha x_i^{(0)})((1-\beta)+\beta \mathbf{\theta}) \\
    ((1-\alpha)\sum_{k=1}^{n} \frac{\hat{A}_{jk}}{\sqrt{\hat{d}_j}\sqrt{\hat{d}_k}}x_k  + \alpha x_j^{(0)})((1-\beta)+\beta \mathbf{\theta}) \\
    \vdots \\
    ((1-\alpha)\sum_{k=1}^{n} \frac{\hat{A}_{nk}}{\sqrt{\hat{d}_n}\sqrt{\hat{d}_k}}x_k  + \alpha x_n^{(0)})((1-\beta)+\beta \mathbf{\theta})
\end{bmatrix}.
\end{align}
}


Consequently,
\small{
\begin{align}
    \mathrm{GNN}(\pi(\begin{bmatrix}
        x_1\\
        x_i\\
        x_j\\
        \vdots \\
        x_n \\
    \end{bmatrix}))& =\begin{bmatrix}
    ((1-\alpha)\sum_{k=1}^{n} \frac{\hat{A}_{1k}}{\sqrt{\hat{d}_1}\sqrt{\hat{d}_k}}x_k  + \alpha x_1^{(0)})((1-\beta)+\beta \mathbf{\theta}) \\
    ((1-\alpha)\sum_{k=1}^{n} \frac{\hat{A}_{jk}}{\sqrt{\hat{d}_j}\sqrt{\hat{d}_k}}x_k  + \alpha x_j^{(0)})((1-\beta)+\beta \mathbf{\theta}) \\
    ((1-\alpha)\sum_{k=1}^{n} \frac{\hat{A}_{ik}}{\sqrt{\hat{d}_i}\sqrt{\hat{d}_k}}x_k  + \alpha x_i^{(0)})((1-\beta)+\beta \mathbf{\theta}) \\
    \vdots \\
    ((1-\alpha)\sum_{k=1}^{n} \frac{\hat{A}_{nk}}{\sqrt{\hat{d}_n}\sqrt{\hat{d}_k}}x_k  + \alpha x_n^{(0)})((1-\beta)+\beta \mathbf{\theta})
\end{bmatrix} \\
& = \pi (\mathrm{GNN}(\begin{bmatrix}
        x_1\\
        x_i\\
        x_j\\
        \vdots \\
        x_n \\
    \end{bmatrix})).
\end{align}
}
From \textbf{Definition 1}, We get
\begin{align}
    \mathrm{GNN}(\pi(\begin{bmatrix}
        x_1\\
        x_i\\
        x_j\\
        \vdots \\
        x_n \\
    \end{bmatrix})) = \pi (\mathrm{GNN}(\begin{bmatrix}
        x_1\\
        x_i\\
        x_j\\
        \vdots \\
        x_n \\
    \end{bmatrix}))=\mathrm{GNN}(\begin{bmatrix}
        x_1\\
        x_i\\
        x_j\\
        \vdots \\
        x_n \\
    \end{bmatrix}).
\end{align}

\begin{flushright}
    Q.E.D
\end{flushright}

\section{D. Reprocessing of HemiBrain dataset}
The HemiBrain dataset~\cite{hemiBrain} used in this study go through several major steps from raw \emph{Drosophila} brains to our model inputs: specimen preparation, imaging, automated segmentation, synapse detection, proofreading, cell type determination, and data reprocessing.

\textbf{Specimen Preparation.}
Starting with a five day old female of wild-type Canton S strain G1 x w1118, the authors use a custom-made jig to microdissect the hemibrain sample with size($250\times250\times250\mathrm{\mu m^3}$), which was then fixed and embedded in Epon, an epoxy resin. Then, the plastic-embedded sample is subdivided into 37 sagittal slabs of 20 µm thickness using~\cite{hot-knife}, with an estimated material loss between consecutive slabs of 30 nm - small enough to allow tracing of long-distance neurites.

\textbf{Imaging.}
 To meet the high isotropic resolution and large volume imaging demand, the authors used FIB-SEM~\cite{FIB} to imaging the slabs, which offers high isotropic resolution (10 nm in x, y, and z), minimal artifacts, and robust image alignment. In order to obtain high-quality slice images faster, the authors use appropriate techniques to solve the problems encountered in practice: \cite{appendix19} for slow imaging speeds; \cite{appendix20} for expanding the practical imaging volume of conventional FIB-SEM; \cite{appendix6} and  \cite{appendix22} for aligning the image stacks, followed by binning along z-axis; \cite{appendix24} for correcting for the warping of the slab, which can occur in that the 20 µm slabs generated by the hot-knife sectioning are embedded in larger plastic tabs prior to FIB-SEM imaging; \cite{appendix25} for adjusting the volumes contrast from deformations introduced by during sectioning, imaging, embedding, and alignment.

\textbf{Automated Segmentation.}
The authors use FFNs~\cite{appendix17} for segmentation and neuronal computational reconstruction of the image data, prior to which CycleGAN \cite{appendix26} is used to adjust the image content to make it more uniform.

\textbf{Synapse Prediction.}
Synapses exist in two forms, namely the presynaptic T-bar and the postsynaptic PSDs, which interconnect to establish communication between two neurons. Initial synaptic predictions suggest the presence of over 9 billion T-bars and 60 billion PSDs in the hemibrain, rendering manual detection prohibitively costly. The authors employ~\cite{appendix29} for synaptic detection, followed by meticulous optimization using~\cite{appendix28}.

\textbf{Proofreading.}
Considering that machine segmentation is not perfect, human proofreading is necessary.
Segmentation errors can be roughly grouped into two classes - false merges, in which two separate neurons are mistakenly merged together, and false splits, in which a single neuron is mistakenly broken into several segments. The authors used~\cite{appendix30} for visualisation and semi-automated proofreading, in which they first addressed large false mergers. In the next phase, the largest remaining pieces are merged into neuron shapes using a combination of machine-suggested edits 31. Finally, the proofreaders used NeuTu~\cite{appendix32} and Neu3~\cite{appendix30} to connect remaining isolated fragments (segments) to already constructed neurons.

\textbf{Cell Type Determination.}
Referring to traditional genetic and optical methods, the authors delineate the neuron type(and sub-cell type, such as layers and slices) of the hemibrain. This process use the criteria of synapse density, glial boundaries, gene expression profile~\cite{appendix33}, neural tracts, and detailed neuron wiring, similar in spirit to the methods of ~\cite{hemiBrain}.

\textbf{Data reprocessing.}
All the above operations are done by the authors of~\cite{hemiBrain}, who put the relevant data on \emph{neuPrint+}(https://neuprint.janelia.org). We download the relevant raw data comprising 21,739 neuronal skeleton datas(3D point sets), which is categorized into 5,555 classes. We subject these categories to a coarsening process, resulting in the consolidation of 5555 initial categories into 380 distinct categories. This consolidation is predicated upon the aggregation of neurons belonging to the same overarching category but distributed across different hierarchical or subcategorical tiers. Subsequent to this treatment, there remain 188 macro-categories encompassing no more than 3 constituent neurons, prompting the exclusion of these instances, amounting to the discarding of 339 neurons.
Among the remaining 192 categories housing 21,400 neurons, these neurons are interlinked by a total of 3,307,248 edges (Connection strength is denoted by the number of synapses between two neurons). Within this assemblage of 21,400 neurons, a subset of 2016 neurons bears the label "None". These neurons are exclusively utilized in training the Connectome Encoder to harness their connectivity; however, during the backpropagation phase for gradient updates, they remain inactive.

\section{E. Reprocessing H01 dataset}
The authors~\cite{h01} acquire a rapidly preserved human surgical sample from the temporal lobe of the cerebral cortex. Similar to HemiBrain, this specimen undergone processes such as sample slicing, imaging, segmentation, among others, culminating in the acquisition of relevant computational reconstructions. The original data is downloaded from the website, comprising 15,554 neurons interlinked by 28,246 edges (edge strength determined by synaptic quantities between two neurons). Notably, 14,458 neurons originate from six distinct layers of the human frontal lobe.  And the remaining 1,096 neurons lack discernible labels; their role is solely confined to supplying connectivity during the training of the Connectome Encoder and does not contribute to backpropagation.

\section{F. Hyperparametric analysis of $\alpha$ and $\beta$ in in main text's Eq. (2).}
We provide more detailed hyper-parameter settings: $\alpha$ in Eq. (2) takes value of 0.1, and $\beta$ in Eq. (2) takes value of 0.5, and the two hyperparameters have been explored experimentally. 

NetNet-C's performance on HemiBrain in different settings of $\alpha$ and $\beta$ are shown in Table~\ref{HemiB_NeuNetC}.
\begin{table}[h]
    \centering
    \resizebox{0.43\linewidth}{!}{
    \begin{tabular}{c|ccccc}
    \toprule
          $\alpha$-$\beta$    & 0.1   & 0.3    & 0.5    & 0.7    & 0.9   \\
    \midrule
          0.1     & 0.5262    & 0.4883    & 0.5168    & 0.5103     & 0.5014  \\
          0.3     & 0.4313 	& 0.4505 	& 0.4280 	& 0.4584 	& 0.4707   \\
          0.5 	& 0.4201 	& 0.3897 	& 0.3617 	& 0.4093 	& 0.3986   \\
          0.7 	& 0.1916 	& 0.1678 	& 0.1921 	& 0.1981 	& 0.1916     \\
          0.9 	& 0.1056 	& 0.1140 	& 0.1308 	& 0.1168 	& 0.1182     \\
    \bottomrule
    \end{tabular}
    }
    \caption{NetNet-C's performance on HemiBrain with varying  $\alpha$ and $\beta$.}
    \label{HemiB_NeuNetC}
\end{table}

NetNet-C's performance on H01 in different settings of $\alpha$ and $\beta$ are shown in Table~\ref{H01_NeuNetC}.
\begin{table}[h]
    \centering
    \resizebox{0.43\linewidth}{!}{
    \begin{tabular}{c|ccccc}
    \toprule
          $\alpha$-$\beta$    & 0.1   & 0.3    & 0.5    & 0.7    & 0.9     \\
    \midrule
        0.1 	& 0.6084 	& 0.6077 	& 0.6141 	& 0.6019 	& 0.6051   \\
        0.3 	& 0.5543 	& 0.5820 	& 0.5884 	& 0.5736 	& 0.5775   \\
        0.5 	& 0.5023 	& 0.4740 	& 0.5138 	& 0.4900 	& 0.5087   \\
        0.7 	& 0.3711 	& 0.3299 	& 0.4013 	& 0.3833 	& 0.3736   \\
        0.9 	& 0.3299 	& 0.3299 	& 0.3299 	& 0.3299 	& 0.3299   \\
    \bottomrule
    \end{tabular}
    }
    \caption{NetNet-C's performance on H01 with varying  $\alpha$ and $\beta$.}
    \label{H01_NeuNetC}
\end{table}

NetNet's performance on HemiBrain in different settings of $\alpha$ and $\beta$ are shown in Table~\ref{HemiBrian_NeuNet}.
\begin{table}[h]
    \centering
    \resizebox{0.43\linewidth}{!}{
    \begin{tabular}{c|ccccc}
    \toprule
          $\alpha$-$\beta$    & 0.1   & 0.3    & 0.5    & 0.7    & 0.9     \\
    \midrule
        0.1 	& 0.9118 	& 0.9061 	& 0.9154 	& 0.9025 	& 0.8762   \\
        0.3 	& 0.9216 	& 0.9174 	& 0.9133 	& 0.9149 	& 0.9138   \\
        0.5 	& 0.9138 	& 0.9195 	& 0.9092 	& 0.9180 	& 0.9128   \\
        0.7 	& 0.9030 	& 0.9112 	& 0.9087 	& 0.9221 	& 0.9102   \\
        0.9 	& 0.9097 	& 0.8978 	& 0.9107 	& 0.9040 	& 0.8968   \\
    \bottomrule
    \end{tabular}
    }
    
    \caption{NetNet's performance on HemiBrain with varying  $\alpha$ and $\beta$.}
    \label{HemiBrian_NeuNet}
\end{table}

NetNet's performance on H01 in different settings of $\alpha$ and $\beta$ are shown in Table~\ref{H01_NeuNet}.
\begin{table}[!h]
    \centering
    \resizebox{0.43\linewidth}{!}{
    \begin{tabular}{c|ccccc}
    \toprule
          $\alpha$-$\beta$    & 0.1   & 0.3    & 0.5    & 0.7    & 0.9     \\
    \midrule
        0.1 	& 0.9448 	& 0.9467 	& 0.9474 	& 0.9488 	& 0.9488   \\
        0.3 	& 0.9448 	& 0.9405 	& 0.9550 	& 0.9439 	& 0.9474   \\
        0.5 	& 0.9343 	& 0.9315 	& 0.9301 	& 0.9377 	& 0.9467   \\
        0.7 	& 0.9426 	& 0.9232 	& 0.9232 	& 0.9100 	& 0.9301   \\
        0.9 	& 0.9059 	& 0.9100 	& 0.9128 	& 0.9163 	& 0.9114   \\
    \bottomrule
    \end{tabular}
    }
    \caption{NetNet's performance on H01 with varying  $\alpha$ and $\beta$.}
    \label{H01_NeuNet}
\end{table}

\section{G. Hyperparametric analysis on number of encoder layers}

\begin{figure}[!h]
	\centering	
	\includegraphics[width=0.80\textwidth]{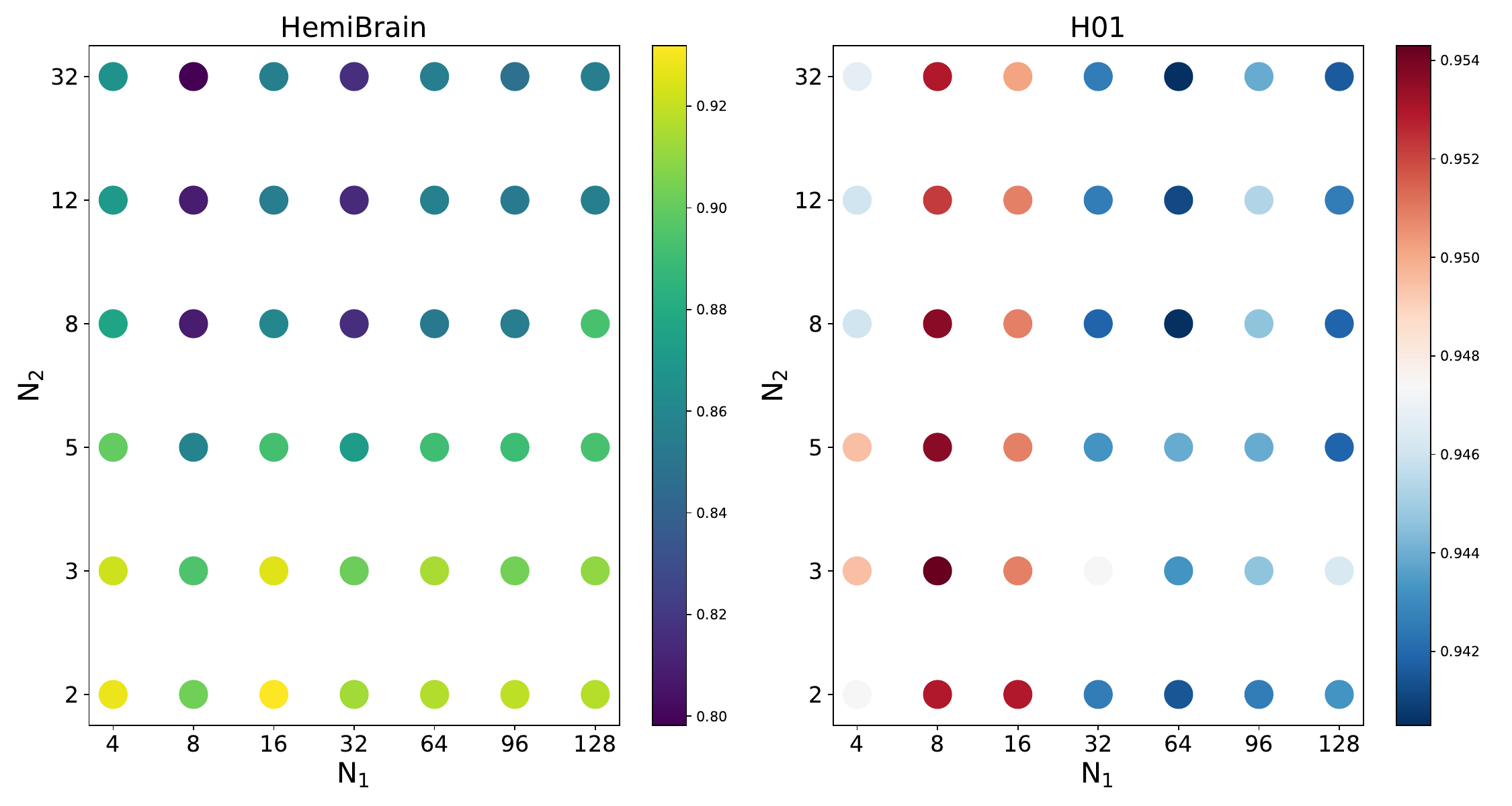}
	\caption{The impact of the number of layers in the Connectome Encoder($N_1$) and Skeleton Encoder($N_2$).}
	\label{num_layer}
\end{figure}

To investigate the impact of the number of layers, $N1$ and $N2$, in  Connectome Encoder and Skeleton Encoder on the experimental results, repectively. We design the variants of NeuNet with different combinations of encoder layers. These variants are employed for neuron classification experiments on the HemiBrain and H01 datasets, and the results are illustrated in Figure \ref{num_layer}. From Figure  \ref{num_layer}, it is evident that the NeuNet demonstrates robustness to variations in the number of layers in both the Connectome Encoder and Skeleton Encoder, consistently maintaining an accuracy above 0.8 across various variants. Considering the trade-off between computational efficiency and accuracy, our final NeuNet configuration employs a 64-layer Skeleton Encoder and a 3-layer Connectome Encoder.

\clearpage
\section{H. Confusion Matrix}
\begin{figure}[h]
    \centering
    
    \subfigure[HemiBrain]{
        \centering
        \includegraphics[width=0.43\textwidth]{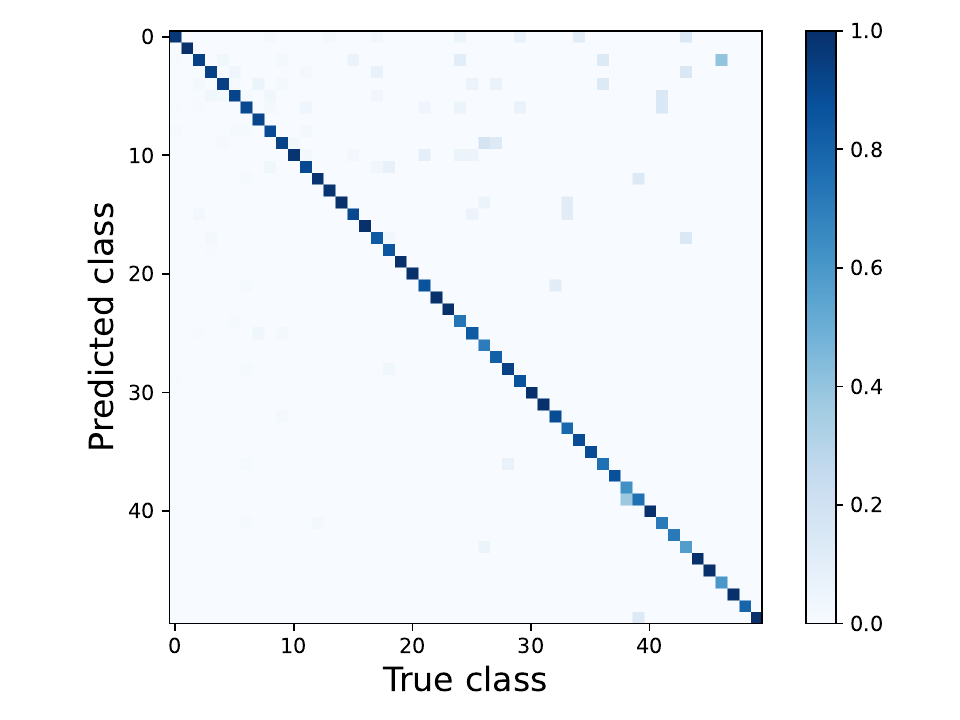}
        \label{confusion_matrix_Hemi}
    }
    \hspace{-15pt}
    \subfigure[H01]{
        \centering
        \includegraphics[width=0.43\textwidth]{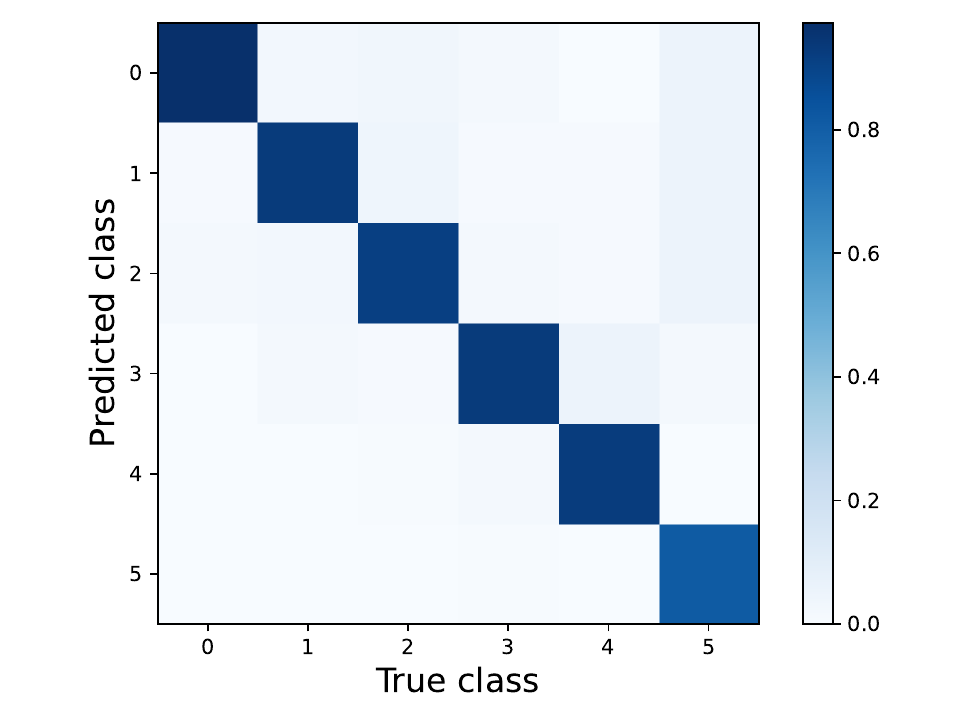}
        \label{confusion_matrix_H01}
    }

    \caption{Accuracy confusion matrix of NeuNet on the HemiBrain and H01 dataset. Where HemiBrain only shows the accuracy of the top-50 classes of 191.}
    \label{confusion_maxtrix}
\end{figure}

The accuracy confusion matrix visualization of the categories for both datasets is shown in Figure~\ref{confusion_maxtrix}, where Figure~\ref{confusion_matrix_H01} shows only the top 50 classes with the highest number. From this, we can see that most of the categories have an accuracy rate of 90\% or higher.

\section{I. HemiBrain's neuron classes}
Delta  (protocerebral bridge Delta between glomeruli) \\
vDeltaA-M  (fan-shaped body vertical Delta within a single column) \\
hDeltaA-M  (fan-shaped body horizontal Delta across columns) \\
EL  (Ellipsoid body - Lateral accessory lobe) \\
EPG  (Ellipsoid body - Protocerebral bridge - Gall) \\
ER  (Ellipsoid body Ring neuron) \\
FB  (Fan-shaped Body) \\
FC1A-3  (Fan-shaped body - Crepine) \\
FR  (Fan-shaped body - Rubus) \\
FS1A-4C  (Fan-shaped body - Superior medial protocerebrum) \\
IbSpsP  (Inferior bridge - Superior posterior slope - Protocerebral bridge) \\
LCNOp  (Lateral accessory lobe - Crepine - NOduli)  \\
LNO  (Lateral accessory lobe - NOduli)  \\
GLNO  (Gall - Lateral accessory lobe - Noduli) \\
LPsP  (Lateral accessory lobe - Posterior slope - Protocerebral bridge) \\
P  (Protocerebral bridge) \\
P6-8P9  (Protocerebral bridge [glomerulus ID1] Protocerebral bridge) \\
PEG  (Protocerebral bridge - Ellipsoid body - Gall) \\
PEN\_a(PEN1),\_b(PEN2)  (Protocerebral bridge - Ellipsoid body - Noduli) \\
PFG  (Protocerebral bridge - Fan-shaped body - Gall surrounding region) \\
PFL (Protocerebral bridge - Fan-shaped body - Lateral accessory lobe) \\
PFN  (Protocerebral bridge - Fan-shaped body - Noduli) \\
PFR  (Protocerebral bridge - Fan-shaped body - Round body) \\
SA1-3  (Superior medial protocerebrum - Asymmetrical body) \\
SAF  (Superior medial protocerebrum - Asymmetrical body - Fan-shaped body) \\
SpsP  (Superior posterior slope - Protocerebral bridge) \\
KC  (Kenyon Cell gamma lobe) \\
MBON0  (Mushroom Body Output Neuron) \\
APL  (Anterior Paired Lateral) \\
DPM  (Dorsal Paired Medial) \\
MB-C  (Mushroom Body - Calyx) \\ 
PAM  (MB-associated DAN, Protocerebral Anterior Medial cluster) \\
PPL  (MB-associated DAN, Protocerebral Posterior Lateral 1 cluster) \\
PPM  (Protocerebral Posterior Medial 1/2 clusters) \\
PAL  (Protocerebral/paired Anterior Lateral cluster) \\
OA-ASM  (OctopAmine - Anterior Superior Medial) \\
OA-VPM  (OctopAmine - ventral paired median) \\
OA-VUM  (OctopAmine - ventral unpaired median anterior) \\
5-HTPLP  (5-HT Posterior lateral protocerebrum) \\
5-HTPMPD  (Posterior medial protocerebrum, dorsal) \\
5-HTPMPV  (Posterior medial protocerebrum, ventral) \\
CSD  (Serotonin-immunoreactive Deutocerebral neuron) \\
AstA1  (Allatostatin A) \\
CRZ  (Corazonin) \\
DSKMP1A, 1B  (Drosulfakinin medial protocerebrum [type ID]) \\
NPFL1-I  (Neuropeptide F lateral large) \\
NPFP  (Neuropeptide F dorso median) \\
PI  (Pars Intercerebralis)  \\
SIF  (SIFamide) \\
DN  (Dorsal Neuron) \\
DN1pA, B  (Dorsal Neuron 1 posterior [type ID]) \\
l-LNv  (large Lateral Neuron ventral) \\
LNd  (Lateral Neuron dorsal) \\
LPN  (Lateral Posterior Neuron) \\
s-LN  (small Lateral Neuron ventral) \\
aDT  (anterior DeuTocerebrum) 

{\setlength{\parindent}{0pt}
aIP  (anterior Inferior Protocerebrum)  \\
aSP-f1-4, g1-3B  (anterior Superior Protocerebrum) \\
aSP8, 10A-10C  (anterior Superior Protocerebrum)  \\
pC  (doublesex-expressing posterior Cells) \\
oviDN  (Oviposition Descending Neuron) \\
oviIN  (Oviposition Inhibitory Neuron) \\
SAG  (Sex peptide Abdominal Ganglion) \\
vpoDN  (vaginal plate opening descending neuron) \\
vpoEN  (vaginal plate opening excitatory neuron) \\
aM  (accessory Medulla) \\
CT  (Complex neuropils Tangential) \\
LC  (Lobula Columnar) \\
LLPC  (Lobula - Lobula Plate Columnar) \\
LPC  (Lobula Plate Columnar) \\
LPLC  (Lobula Plate - Lobula Columnar) \\
LT  (Lobula Tangential) \\
MC  (Medulla Columnar) \\
DCH  (Dorsal Centrifugal Horizontal) \\
H  (Horizontal) \\
HSN, E, S  (Horizontal System North, Equatorial, South) \\
VS  (Vertical System)
VCH  (Ventral Centrifugal Horizontal) \\
Li  (Lobula intrinsic) \\
HB  (Hofbauer-Buchner)  \\
DN  (Descending Neuron cell) \\
DNES1-3  (Descending Neuron going out to ESophagus) \\
MDN  (Moonwalker Descending Neuron) \\
ORN\_D, DA, DC, DL, DM, DP, VA, VC, VL, VM  (Olfactory Receptor Neuron) \\
TRN\_VP  (Thermo-Receptor Neuron) \\
HRN\_VP  (Hygro-Receptor Neuron) \\
JO-ABC  (Johnston’s Organ auditory receptor neuron- [AMMC zone ID]) \\
OCG  (OCellar Ganglion neuron) \\
D\_adPN, DA1\_lPN, DC2\_adPN, DL3\_lPN, DM4\_vPN, DP1l\_adPN, VA1d\_adPN, VC2\_lPN, VL2p\_vPN, VM7d\_adPN, VP2\_l2PN  (uniglomerular [glomerulus ID] \_ [cell cluster ID] Projection Neuron) \\
VP1l+\_lvPN, VP3+\_vPN  (uni+glomerular [glomerulus ID]+ \_ [cell cluster ID] Projection Neuron, arborizing in a glomerulus and a few neighboring areas) \\
VP1m+VP2\_lvPN, VP4+VL1\_l2PN  (biglomerular [glomerulus ID1], \_ [cell cluster ID] Projection Neuron, arborizing in two glomeruli) \\
M\_smPN, adPN, spPN, lPNm11A-13, l2PN, lvPNm24-48, lv2PN, vPN, ilPN, imPN   (Multiglomerular\_ [cell cluster ID] Projection Neuron) \\
MZ\_lvPN, lv2PN  (Multiglomerular and subesophageal Zone \_ [cell cluster ID] Projection Neuron) \\
Z\_lvPN, Z\_vPN  (subesophageal Zone only \_ [cell cluster ID] Projection Neuron ) \\
lLN, v2LN, il3LN, l2LN, vLN  ([cell cluster ID] Local Neuron ) \\
mAL, B, C, D  (mediodorsal Antennal Lobe neuron [type ID]) \\
AL-AST  (Antennal Lobe - Antenno-Subesophageal Tract) \\
AL-MBDL  (Antennal Lobe - Median BunDLe) \\
ALBN  (Antennal Lobe Bilateral Neuron) \\
LHAD  (Lateral Horn Anterior Dorsal cell cluster) \\
ALIN  (Antennal Lobe INput neuron) \\
LHAV  (Lateral Horn Anterior Ventral cell cluster) \\
LHPD  (Lateral Horn Posterior Dorsal cell cluster) \\
LHPV  (Lateral Horn Posterior Ventral cell cluster) \\
LHCENT  (Lateral Horn CENTrifugal) \\
LHMB  (Lateral Horn - Mushroom Body) \\
AOTU  (Anterior Optic TUbercle) \\
TuBu01-10, A, B (anterior optic Tubercle - Bulb [type ID]) \\
ATL  (Antler) \\
AVLP  (Anterior VentroLateral Protocerebrum) \\
CL  (CLamp) \\
CRE  (CREpine) \\
IB  (Inferior Bridge) \\
LAL  (Lateral Accessory Lobe) \\
PLP  (Posterior Lateral Protocerebrum) \\
PS  (Posterior Slope) \\
PVLP  (Posterior VentroLateral Protocerebrum) \\
SAD  (SADdle) \\
AMMC-A  (Antennal Mechanosensory and Motor Center) \\
SLP  (Superior Lateral Protocerebrum) \\
SIP  (Superior Intermediate Protocerebrum) \\
SMP  (Superior Medial Protocerebrum) \\
DGI  (Dorsal Giant Interneuron) \\
VES  (VESt) \\
WED  (WEDge) \\
WEDPN  (WEDge Projection Neuron) }

\twocolumn
\section{Acknowledgments}
This work is partially supported by National Key R\&D Program of China (2023YFC2705700), the National Natural Science Foundation of China (62206202, 62225113), China Postdoctoral Science Foundation, China (2022M712461), Artificial Intelligence Innovation Project of Wuhan Science and Technology Bureau (No. 2022010702040070), The Fundamental Research Funds for the Central Universities, China (2042022kf1043).

\bibliography{aaai24}


\end{document}